\documentclass[journal=jacsat,manuscript=article]{achemso}
\usepackage{mciteplus}

\usepackage{tikz}
\usetikzlibrary{arrows.meta, positioning, shapes.geometric}
\usepackage[utf8]{inputenc}    
\usepackage[T1]{fontenc}       
\usepackage{lmodern}           
\usepackage{textcomp}
\usepackage[utf8]{inputenc}
\usepackage{braket} 
\usepackage{amsmath, amssymb} 
\usepackage{tcolorbox}
\tcbuselibrary{theorems, skins, breakable}

\usepackage[version=3]{mhchem} 

\usepackage{newunicodechar}
\newunicodechar{−}{-}

\author{Sk Mujaffar Hossain}
\affiliation[IKST]
{Indo-Korea Science and Technology Center (IKST), Bangalore, India}

\author{Seung-Cheol Lee}
\email{leesc@kist.re.kr}
\affiliation[IKST]
{Indo-Korea Science and Technology Center (IKST), Bangalore, India}
\alsoaffiliation[EMRC KIST]
{Electronic Materials Research Center, Korea Institute of Science $\&$ Technology, Korea}

\author{Satadeep Bhattacharjee}
\email{s.bhattacharjee@ikst.res.in}
\affiliation[IKST]
{Indo-Korea Science and Technology Center (IKST), Bangalore, India}

\title[\textsf{achemso}]
    {Quantum Simulations of Battery Electrolytes with VQE–qEOM and SQD: Active-Space Design, Dissociation, and Excited States of LiPF$_6$, NaPF$_6$, and FSI Salts}

\abbreviations{IR,NMR,UV}

\begin{document}



\begin{abstract}
Accurate prediction of excited states in battery electrolytes is central to understanding photostability, oxidative stability, and degradation. We employ hybrid quantum--classical algorithms---the Variational Quantum Eigensolver (VQE) for ground states combined with the quantum equation of motion (qEOM) for vertical singlet excitations---to study LiPF$_6$, NaPF$_6$, LiFSI, and NaFSI. Compact active spaces were constructed from frontier orbitals, mapped to qubits, and reduced via symmetry tapering and commuting-group measurements to lower sampling cost. Within $\sim$10-qubit models, VQE--qEOM agrees closely with exact diagonalization of the same Hamiltonians, while sample-based quantum diagonalization (SQD) in larger active spaces recovers near-exact (subspace-FCI) energies. The spectra display clear anion and cation trends: PF$_6$ salts exhibit higher first-excitation energies (e.g., LiPF$_6$ $\approx$13.2~eV) and a compact three-state cluster at 12--13~eV, whereas FSI salts show substantially lower onsets ($\approx$8--9~eV) with a near-degenerate (S$_1$,S$_2$) followed by S$_3$ $\sim$1.3~eV higher. Substituting Li$^+$ with Na$^+$ narrows the gap by $\sim$0.4--0.8~eV within each anion family. Converting S$_1$ to wavelengths places the onsets in the deep-UV (LiPF$_6$~$\sim$94~nm; NaPF$_6$~$\sim$100~nm; LiFSI~$\sim$141~nm; NaFSI~$\sim$148~nm). All results pertain to isolated species or embedded clusters appropriate to the NISQ regime; solvent shifts can be incorporated a posteriori via classical $\Delta$-solvation or static embedding. These results demonstrate that current quantum algorithms can deliver chemically meaningful excitation and binding trends for realistic electrolyte motifs and provide quantitative baselines to guide electrolyte screening and design.
\end{abstract}

\section{Introduction}

The rapid development of electrochemical energy storage technologies has created an urgent demand for stable, high-performance electrolytes. In lithium-ion batteries (LIBs), and increasingly in sodium-ion batteries (SIBs), electrolyte salts such as LiPF$_6$, NaPF$_6$, LiFSI, and NaFSI \cite{he2025research} play a pivotal role in determining ionic conductivity, oxidative stability, and long-term durability. However, the excited-state properties of these salts remain poorly understood, despite their direct relevance to photostability, oxidative degradation, and safety under extreme conditions. Classical approaches such as density functional theory (DFT) and post-Hartree--Fock methods have provided valuable insights, yet they face well-known limitations in treating electron correlation and excited states in complex molecular systems. 


Recent advances in quantum computing provide a promising alternative by enabling direct simulation of many-body electronic structures on qubit-based devices\cite{patel2025quantum,cao2019quantum}. Hybrid quantum--classical algorithms such as the Variational Quantum Eigensolver (VQE)\cite{peruzzo2014variational,boyn2021quantum, mcclean2016theory} and its extension with the quantum Equation-of-Motion (qEOM)\cite{zheng2024quantum,hlatshwayo2024quantum} offer scalable strategies for accessing both ground- and excited-state energies within the noisy intermediate-scale quantum (NISQ) regime \cite{peruzzo2014variational,mcclean2016theory,ollitrault2020quantum,de2023complete,bharti2022noisy,acampora2023d}. Complementary to VQE--qEOM, sample-based quantum diagonalization (SQD)\cite{robledo2025chemistry,yu2025quantum,kanno2023quantum,sugisaki2024hamiltonian} provides a systematic route to recover nearly exact eigenvalues in larger active spaces, thereby extending the accuracy and benchmarking capability of quantum simulations. Together, these methods open new possibilities for probing photo-induced processes, excitation spectra, and stability trends in battery electrolytes beyond the reach of classical approaches.

Recent efforts have applied quantum algorithms to chemical and materials problems, demonstrating their potential but also highlighting key challenges\cite{patel2025quantum,sureshbabu2021implementation}. The VQE has been successfully employed for small molecules as toy model such as H$_2$, LiH, BeH$_2$, H$_4$, HeH$^+$, and CO$_2$\cite{peruzzo2014variational,mcclean2016theory,o2016scalable,ghosh2023deep, tao2022exploring,lotstedt2022evaluation,bauer2020quantum}, and extended to larger systems using symmetry-based qubit reduction and tailored ans\"atze \cite{cao2019quantum}. Beyond all these studies, quantum computations are increasingly being explored in diverse fields such as drug discovery \cite{cao2019quantum,bauer2020quantum, blunt2022perspective,cava2021introduction,delgado2025quantum}, sustainable catalysis \cite{hariharan2024modeling,lipka2024catalysis,von2021quantum}, organic photovoltaics \cite{ollitrault2020quantum}, strongly correlated quantum materials \cite{motta2020quantum}, electronic structure\cite{barkoutsos2018quantum}, and nonadiabatic molecular quantum dynamics\cite{ollitrault2020nonadiabatic,miessen2021quantum}. These applications demonstrate the broad potential of quantum algorithms to address problems across chemistry, materials science, and biology, further motivating their deployment in complex molecular systems such as battery electrolytes. These advances highlight the versatility of quantum approaches across diverse domains of chemistry, from biomolecular design to energy materials\cite{paudel2022quantum}.
 The qEOM approach has further enabled access to electronically excited states, providing chemically accurate excitation energies in benchmark systems \cite{ollitrault2020quantum,mcardle2020quantum}. 

More recently, applications have begun to emerge in energy materials\cite{paudel2022quantum}, including quantum simulations of sulfur species in lithium--sulfur batteries \cite{rice2021quantum}, organic emitters \cite{gao2021applications}, and transition-metal complexes \cite{ghosh2023deep}. Despite these advances, systematic studies of realistic electrolyte salts, particularly those relevant for Li- and Na-ion batteries, remain unexplored. In this context, our work establishes a framework that integrates VQE, qEOM, and SQD to capture both ground- and excited-state properties of complex electrolyte systems with chemical accuracy, while revealing chemically meaningful trends in photostability and oxidative stability.

In this work, we apply VQE + qEOM to investigate the ground and low-lying excited states of four representative electrolyte salts: LiPF$_6$, NaPF$_6$, LiFSI, and NaFSI. By systematically designing active spaces, reducing qubit requirements, and benchmarking against classical solvers, we establish chemically accurate simulations within compact quantum models and validate scalability using SQD. Beyond excitation energies, we analyze HOMO--LUMO gaps, dissociation profiles, and photophysical pathways relevant to electrolyte degradation. Our results reveal clear trends between PF$_6^-$ and FSI$^-$ salts as well as Li$^+$/Na$^+$ analogues, offering chemically meaningful insights into photostability and oxidative stability. This work demonstrates the feasibility of quantum algorithms for realistic electrolyte systems and marks a significant step toward leveraging quantum simulations for the rational design of next-generation battery materials.All quantum results reported here pertain to isolated species or embedded clusters compatible with NISQ resources. 

\section{Methodology}

\subsection{Workflow of Quantum Simulation}
\begin{figure}[htp]
    \centering
    \includegraphics[width=16.5cm]{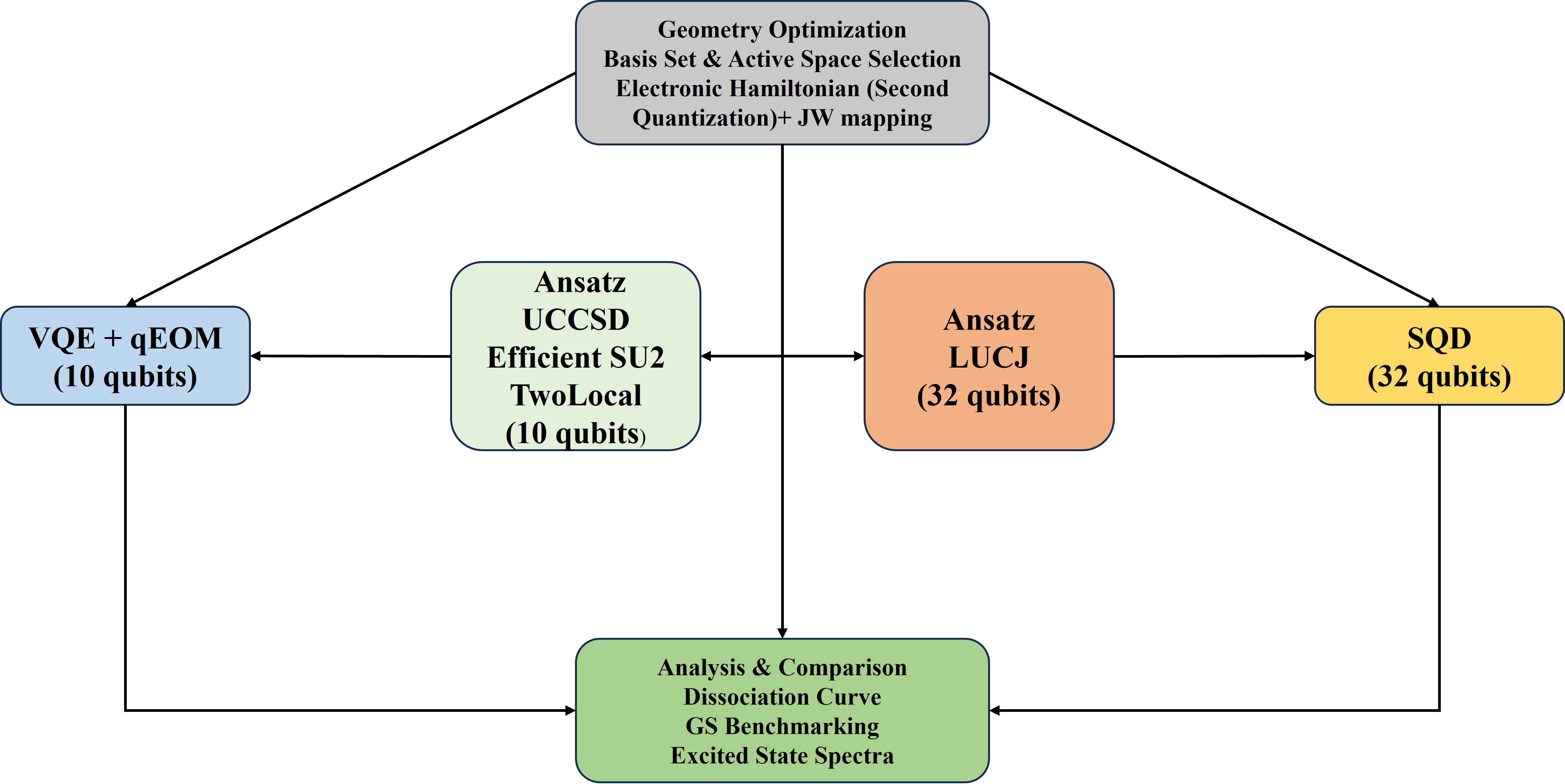}
    \caption{Workflow of quantum simulation study of molecular electrolytes.}
    \label{fig:workflow}
\end{figure}

In this study, we investigated the ground and excited state properties of representative electrolyte salts, including LiPF$_6$, NaPF$_6$, LiFSI, and NaFSI, using hybrid quantum--classical algorithms. The workflow in Figure \ref{fig:workflow} consisted of (i) construction of active spaces from \textit{ab initio} orbital energies, (ii) mapping of the molecular Hamiltonian to qubit operators, (iii) ground-state optimization using the Variational Quantum Eigensolver (VQE)\cite{bauer2020quantum}, (iv) excited-state calculations using the quantum Equation-of-Motion (qEOM) formalism, and (v) benchmarking and scaling analysis using Sample-based Quantum Diagonalization (SQD). All quantum simulations were implemented using IBM Qiskit \cite{matthew_treinish_2023_8190968}, which was executed primarily on noiseless state vector simulators, later executed on real quantum hardware computers.

\subsection{Geometry Optimization, Basis Set Choice, and Active Space Design}

The molecular geometries of LiPF$_6$, NaPF$_6$, LiFSI, and NaFSI were optimized at the B3LYP/6-311+G(d,p) level of theory using the Gaussian software package\cite{g16}. For quantum computations, several basis sets (STO-3G, 6-31G, and cc-pVDZ) were benchmarked to balance computational cost and accuracy. The cc-pVDZ basis set was ultimately selected for production runs, as it most widely used \cite{kaliakin2025implicit} and provides a reliable compromise between capturing electronic correlation and maintaining tractable quantum resource requirements.  

Active spaces were systematically constructed by freezing chemically inert core orbitals and retaining only the valence orbitals near the HOMO–LUMO frontier. This strategy reduced the number of required qubits from more than 200 in the full orbital space to a practical range of 4 to 12 qubits, while preserving the essential correlation effects required for chemical accuracy. The active-space choices were further validated by inspecting molecular orbital visualizations and natural occupation numbers, ensuring that the reduced spaces retained all electronically relevant contributions. This protocol follows best practices established in recent active space design studies for quantum simulations of the NISQ era \cite{motta2020quantum, tilly2022variational}.

\subsection{Electronic Hamiltonian and Qubit Mapping}

The molecular ground state was obtained through the Rayleigh--Ritz variational principle, which guarantees the following.
\begin{equation}
\langle \Psi(\vec{\Theta}) \,|\, \hat{H} \,|\, \Psi(\vec{\Theta}) \rangle \geq E_{0},
\end{equation}
where $\Psi(\vec{\Theta})$ is a parametrized trial wavefunction, $\hat{H}$ is the molecular Hamiltonian operator, and $E_{0}$ is the exact ground state energy. Within the VQE framework, the parameters $\vec{\Theta}$ are variationally optimized such that the expectation value approaches $E_{0}$ as closely as possible on quantum hardware or simulators.

The molecular electronic Hamiltonian $\hat{H}$ was expressed in second quantization within the Born--Oppenheimer approximation as\cite{de2023complete,ghosh2023deep,gao2021applications}
\begin{equation}
\hat{H} = \sum_{pq} h_{pq} \, a_p^\dagger a_q 
+ \tfrac{1}{2} \sum_{pqrs} h_{pqrs} \, a_p^\dagger a_q^\dagger a_r a_s
\end{equation}
where $a_p^\dagger$ and $a_q$ are fermionic creation and annihilation operators, $h_{pq}$ are one-electron integrals (kinetic energy and nuclear attraction), and $h_{pqrs}$ are two-electron repulsion integrals in an orthonormal molecular orbital basis and these one and two-electron integrals were obtained using the PySCF package.\cite{matthew_treinish_2023_8190968,sun2020recent,sun2018pyscf,ekström2010arbitrary}
For quantum simulation, the fermionic Hamiltonian was mapped to qubit operators using the Jordan--Wigner (JW)\cite{jordan1928pauli,ortiz2002simulating,somma2002simulating} transformation, which preserves the fermionic anticommutation relations by introducing non-local strings of Pauli operators. The resulting qubit Hamiltonian takes the generic form
\begin{equation}
\hat{H} = \sum_i c_i P_i = \sum_i c_i\prod_j\sigma_j^i 
\end{equation}
where $c_i$ are real coefficients derived from molecular integrals and $P_i$ are tensor products of Pauli operators $\sigma$ ($I$, $X$, $Y$, $Z$)\cite{de2023complete}. This qubit representation serves as the input for the VQE and subsequent qEOM excited-state calculations.

\subsection{Variational Quantum Eigensolver (VQE)}
The VQE algorithm was used to obtain ground-state wavefunctions by variationally minimizing the expectation value of the molecular Hamiltonian on a quantum device or simulator.\cite{peruzzo2014variational,mcclean2016theory,grimsley2019adaptive} 
Fermionic Hamiltonians were mapped to qubit operators via the Jordan--Wigner transformation, and several ans\"atze were benchmarked, including hardware-efficient forms (EfficientSU2, TwoLocal) and the chemically motivated UCCSD.\cite{peruzzo2014variational,romero2018strategies,grimsley2019trotterized,evangelista2019exact,kandala2017hardware} 
Although UCCSD entails greater circuit depth and is less NISQ-friendly, it consistently delivered the highest accuracy for electron correlation, in line with prior studies.\cite{innan2024quantum}

Classical optimization employed gradient-free COBYLA\cite{powell1994direct} (maximum 1000 iterations). 
Convergence was monitored using both the energy change (threshold $\Delta E \le 10^{-9}$~Ha) and the energy variance. 
The optimized ground state $|\psi_0\rangle$ served as the reference for all qEOM excited-state calculations. To reduce measurement cost, Pauli terms were grouped by qubit-wise commutativity. 
Unless otherwise noted, typical budgets were $5{\times}10^3$--$2{\times}10^4$ shots per geometry point under the same $\Delta E \le 10^{-9}$~Ha convergence criterion.

\subsection{Quantum Equation-of-Motion (qEOM)}

To investigate electronically excited states, we employed the quantum equation-of-motion (qEOM) formalism, which builds upon the VQE-optimized ground state reference by invoking a linear response framework \cite{ollitrault2020quantum,santagati2018witnessing,higgott2019variational,mcclean2017hybrid,colless2018computation}. In this method, excitation operators $\hat{R}_i$ are constructed within the particle-conserving subspace, typically including single and double excitations relative to the ground state. The excited-state ansatz takes the form
\begin{equation}
|\psi_i\rangle = \hat{R}_i |\psi_0\rangle ,
\end{equation}
where $|\psi_0\rangle$ is the VQE ground state and $\hat{R}_i$ acts as an excitation operator.  

The excitation energies $\omega_i$ are obtained by solving the eigenvalue equation
\begin{equation}
[\hat{H}, \hat{R}_i] |\psi_0\rangle = \omega_i \hat{R}_i |\psi_0\rangle ,
\end{equation}
which, in practice, is recast into a generalized eigenvalue problem by evaluating the commutator matrix elements on a quantum device or simulator. This procedure yields the low-lying vertical excitation spectrum directly from the ground-state reference, without the need for separate state-specific variational optimizations.  

The qEOM approach has been demonstrated to provide chemically accurate excitation energies for small and medium-sized molecules \cite{mcardle2020quantum,fan2021equation}, and is particularly attractive in the NISQ era because it leverages the compact ground-state wavefunction from VQE while accessing a manifold of excited states at comparable cost. In the context of battery electrolytes, qEOM enables characterization of photo-induced excitations and charge-transfer processes that are crucial for understanding stability and degradation mechanisms under operating conditions.

\subsection{Sample-Based Quantum Diagonalization (SQD)}

While VQE combined with qEOM provided chemically accurate results within reduced active spaces (10 qubits), we further employed Sample-Based Quantum Diagonalization (SQD)\cite{robledo2025chemistry,nakagawa2024adapt,kanno2023quantum,barison2025quantum} to benchmark accuracy in significantly larger orbital subspaces. SQD is a projector-based hybrid quantum--classical algorithm that constructs an effective Hamiltonian by stochastically sampling matrix elements of $\hat{H}$ on a quantum device, followed by classical diagonalization \cite{huggins2022unbiasing,lee2022unbiasing}. This approach offers two key advantages. First, by decoupling Hamiltonian sampling from variational optimization, SQD avoids barren-plateau issues and optimizer instabilities that limit VQE scalability. Second, SQD naturally accommodates larger active spaces---here extended up to 32 qubits---allowing systematic recovery of correlation energy beyond the reach of compact variational ans\"atze.  

In addition to UCCSD, we also employed the Local Unitary Cluster Jastrow (LUCJ) ansatz \cite{motta2020quantum}. While UCCSD is chemically motivated and excels at dynamic correlation, it suffers from excessive circuit depth. By contrast, LUCJ offers hardware efficiency by capturing short- and medium-range correlations with reduced depth, making it more suitable for NISQ devices. By combining SQD with LUCJ, we achieved robust ground-state energies in large active spaces, thereby highlighting the complementary strengths of variational ans\"atze and projector-based diagonalization. Consistent with recent findings \cite{kaliakin2025implicit}, SQD yielded unbiased eigenvalues with polynomial sampling cost, validated the reduced-qubit VQE results, and demonstrated near-exact recovery of correlation energy as the active space was enlarged.  

\subsubsection*{Mathematical Formulation of SQD}

Sample-Based Quantum Diagonalization (SQD) is a projector-based hybrid quantum--classical approach that reconstructs approximate eigenstates of the molecular Hamiltonian using information obtained directly from quantum measurements. Instead of optimizing a parameterized wavefunction, as in VQE, SQD works by sampling and post-processing measurement outcomes, followed by classical diagonalization in a reduced subspace. This provides a scalable route for accessing ground and excited states without relying on deep parameterized circuits. The procedure can be summarized in four main steps:

\textbf{1. Configuration recovery.}  
Quantum measurements on $n$-qubit states yield bitstrings $\{x\}$ that encode orbital occupation patterns. However, due to noise or finite sampling, many bitstrings may violate conserved physical symmetries such as particle number or total spin. SQD corrects such errors by probabilistically flipping occupations so that each corrected bitstring $\tilde{x}$ satisfies the target constraints (e.g., fixed electron number). This ensures that the working distribution of configurations $\{\tilde{x}\}$ is consistent with the physical Hilbert space of the molecule.

\textbf{2. Subsampling.}  
From the corrected pool $\{\tilde{x}\}$, a representative subset of $N_s$ bitstrings is selected. Each bitstring $\ket{x_i}$ defines a computational basis vector, and together they span a reduced working subspace 
\[
S = \mathrm{span}\{\ket{x_1}, \ket{x_2}, \ldots, \ket{x_{N_s}}\}.
\]
The choice of $N_s$ controls the trade-off between accuracy and computational cost. For large systems, this step can be parallelized by dividing samples into independent batches, each yielding a local approximation to the target eigenstate.

\textbf{3. Subspace diagonalization.}  
Once the subspace $S$ is defined, the algorithm constructs a projected Hamiltonian
\[
H_S = P_S \hat{H} P_S, \quad P_S = \sum_{i=1}^{N_s} \ket{x_i}\bra{x_i},
\]
with matrix elements
\[
(H_S)_{ij} = \bra{x_i}\hat{H}\ket{x_j}, \quad (S)_{ij} = \braket{x_i|x_j}.
\]
The pair $(H_S, S)$ defines a generalized eigenvalue problem
\[
H_S \mathbf{c} = E \, S \mathbf{c},
\]
which can be solved classically to yield approximate eigenpairs
\[
\big(E, \; \ket{\Psi} = \sum_{i=1}^{N_s} c_i \ket{x_i}\big).
\]
This diagonalization recovers both ground and excited states simultaneously as the lowest and higher eigenroots.

\textbf{4. Iteration.}  
The lowest-energy eigenstate $\ket{\Psi}$ obtained from diagonalization is used to update orbital occupancies, which are fed back into the configuration recovery step. This iterative loop is repeated until self-consistency is reached, ensuring that the sampled subspace converges toward the true eigenstate.

The efficiency of SQD arises from the fact that many chemically relevant eigenstates are \emph{sparse} in the computational basis, meaning that they can be faithfully approximated by a relatively small number $N_s$ of bitstrings. Thus, SQD reduces the exponentially large $2^n \times 2^n$ Hamiltonian to a tractable $N_s \times N_s$ matrix with $N_s \ll 2^n$. Furthermore, because excited states are obtained as higher eigenroots, SQD naturally provides access to optical excitations and transition dipoles. From these, oscillator strengths can be evaluated, enabling direct photophysical and stability analyses of electrolyte molecules.



\subsection{Simulation Details}

For each electrolyte molecule, both ground- and excited-state simulations were performed across multiple active spaces and ans\"atz choices. Dissociation energy profiles were obtained by systematically scanning cation–anion distances and computing the total energies at each geometry. Electronic properties, including HOMO–LUMO gaps and excitation thresholds, were extracted from the qEOM spectra. All quantum results were benchmarked against classical exact diagonalization (CASCI) within the chosen same active spaces to assess accuracy. The performance of each approach was analyzed in terms of energy deviations, convergence behavior, and photophysical trends relevant to electrolyte degradation mechanisms. To ensure hardware compatibility, all ans\"atze (UCCSD and LUCJ) were transpiled into optimized quantum circuits using the Qiskit transpiler. Simulations were executed using the IBM Qiskit Runtime service\cite{matthew_treinish_2023_8190968,nation2023suppressing}, with selected runs carried out on the \textit{ibm\_brisbane} quantum device, thereby providing both noiseless simulator benchmarks and hardware-level validation. All VQE–qEOM and SQD results reported here are obtained for isolated species or embedded cluster Hamiltonians that do not explicitly model the macroscopic solvent.

\section{Results and discussion}
While classical quantum chemistry methods continue to be the gold standard in terms of accuracy and computational maturity, they face significant scalability challenges when modeling increasingly complex and strongly correlated molecular systems. In contrast, quantum algorithms such as the Variational Quantum Eigensolver (VQE) and quantum Equation of Motion (qEOM) present a promising alternative, particularly in the context of the Noisy Intermediate-Scale Quantum (NISQ) era.

In this study, we have employed these quantum algorithms not only to assess their computational performance and resource efficiency, but also to critically evaluate their scalability and practical applicability to chemically relevant systems such as battery electrolyte molecules. Our findings provide a foundational benchmark for the quantum simulation of complex molecular systems and demonstrate effective strategies for qubit optimization—an essential requirement for extending quantum computations to larger, industrially significant materials.

This work thus serves as a stepping stone toward practical quantum computational chemistry, charting a roadmap for future large-scale simulations of energy materials in both the NISQ and fault-tolerant quantum computing regimes.

\subsection{Design of Active Spaces}

\begin{figure}[htp]
    \centering
    \includegraphics[width=15cm]{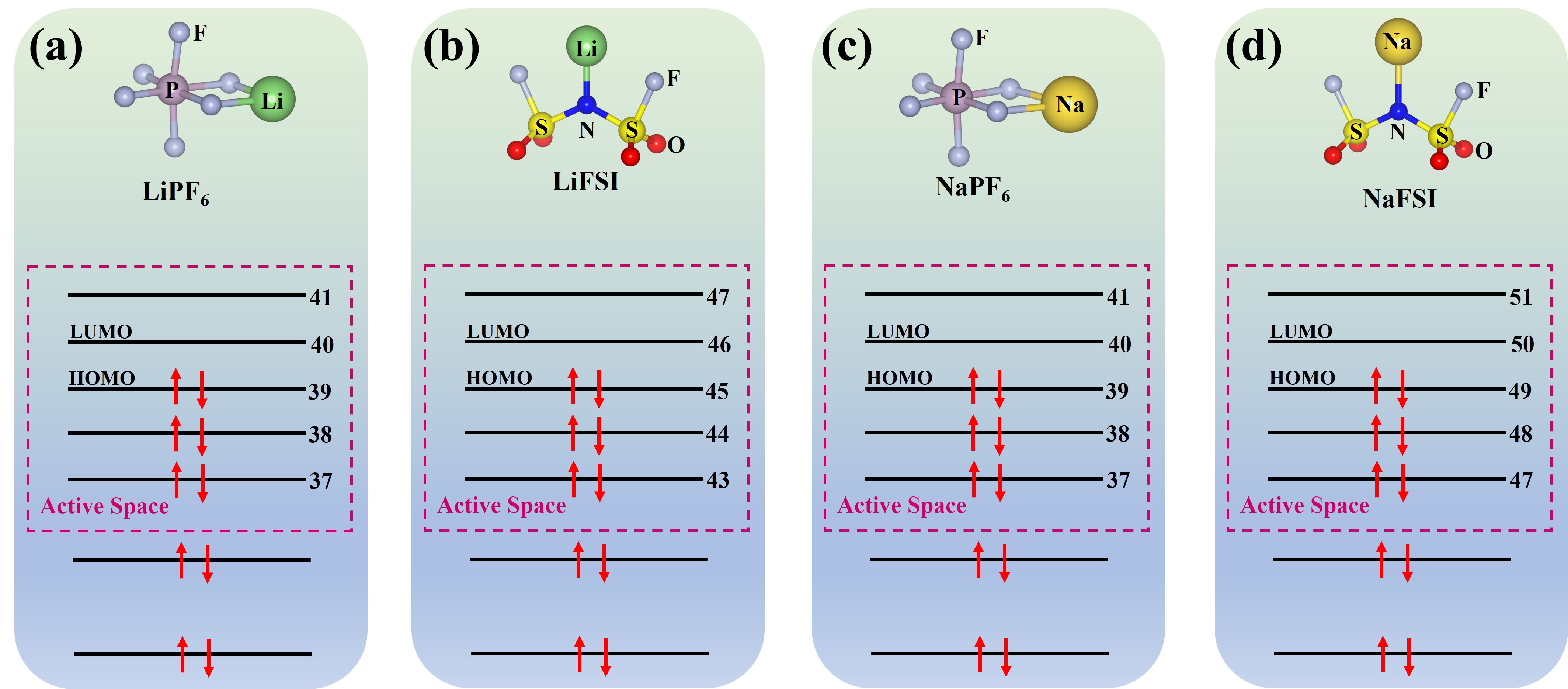}
    \caption{Active spaces selected for (a) LiPF$_6$, (b) LiFSI, (c) NaPF$_6$, and (d) NaFSI, showing molecular structures and corresponding frontier orbitals. Red arrows indicate the orbitals included in the active space, chosen near the HOMO–LUMO gap to capture essential electronic excitations for quantum simulations.}
    \label{fig:active_space}
\end{figure}

Due to the relatively large size and chemical complexity of the electrolyte molecules considered in this study: LiPF$_6$, NaPF$_6$, LiFSI and NaFSI, it is computationally intractable to simulate the full orbital space within the current limitations of quantum hardware. Therefore, an appropriate choice of active space is essential to balance accuracy and resource efficiency. Active space is constructed by selecting a subset of molecular orbitals that contribute significantly to the ground and excited states, typically those surrounding the HOMO–LUMO region. This selection ensures that essential electron correlation and excitation effects are captured while minimizing the total number of qubits required (see Table \ref{tab:qubit_reduction}). Figure~\ref{fig:active_space} illustrates the active spaces used for each molecule, highlighting the specific occupied and virtual orbitals retained for quantum simulation. The selected active spaces were validated by inspecting the numbers of natural orbital occupation and visualizing key molecular orbitals to ensure physical relevance for both ground and excited-state properties.

\begin{figure}[htp]
    \centering
    \includegraphics[width=15cm]{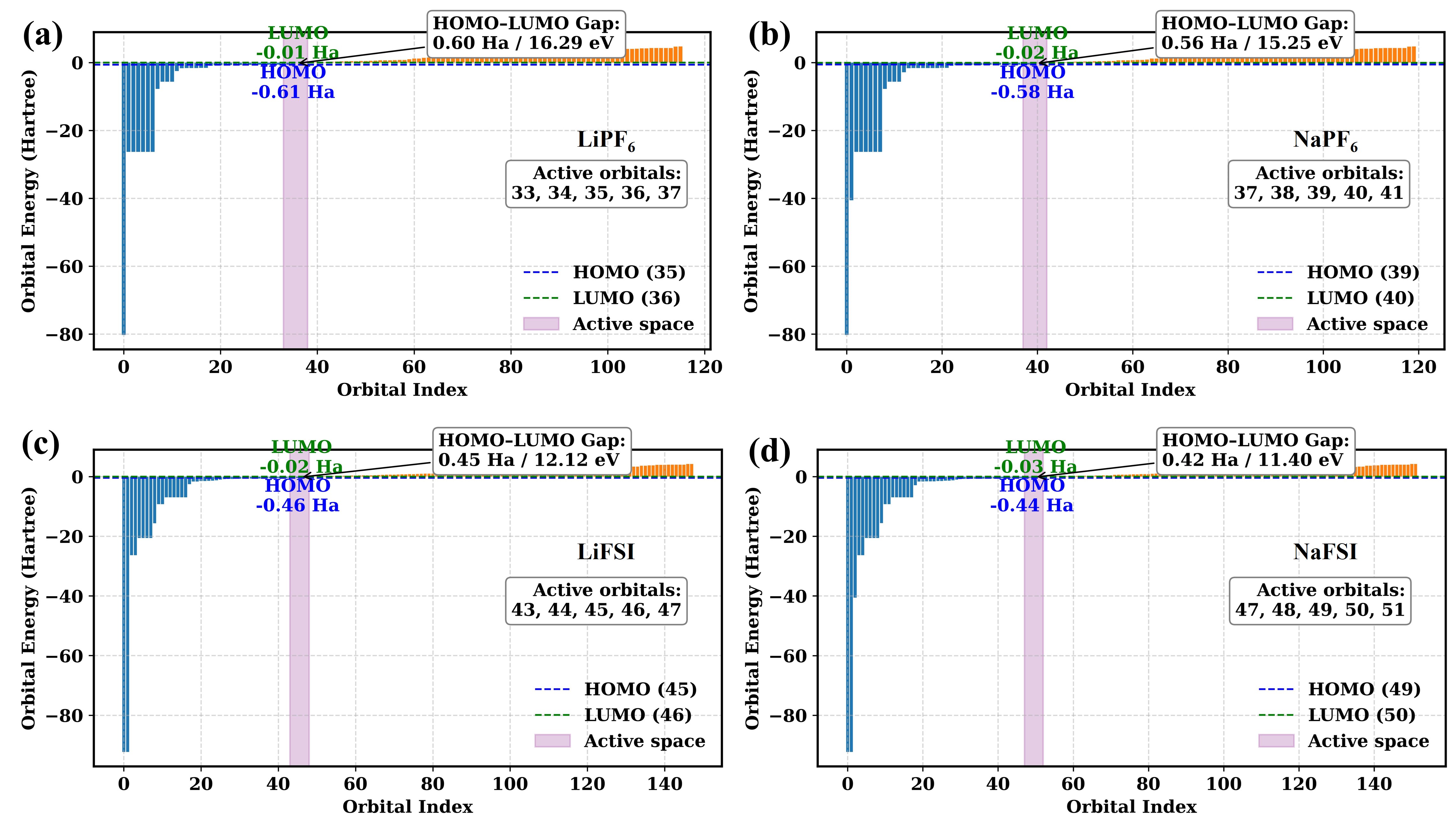}
    \caption{Active spaces selected for (a) LiPF$_6$, (b) LiFSI, (c) NaPF$_6$, and (d) NaFSI.}
    \label{fig:active_orbital}
\end{figure}

Figure \ref{fig:active_orbital} represents the specific region close to the HOMO and LUMO with a particular list of orbitals considered for active space during the quantum simulation of each molecules.

\subsection{Qubit Reduction and Optimization}
\begin{figure}[htp]
    \centering
    \includegraphics[width=12cm]{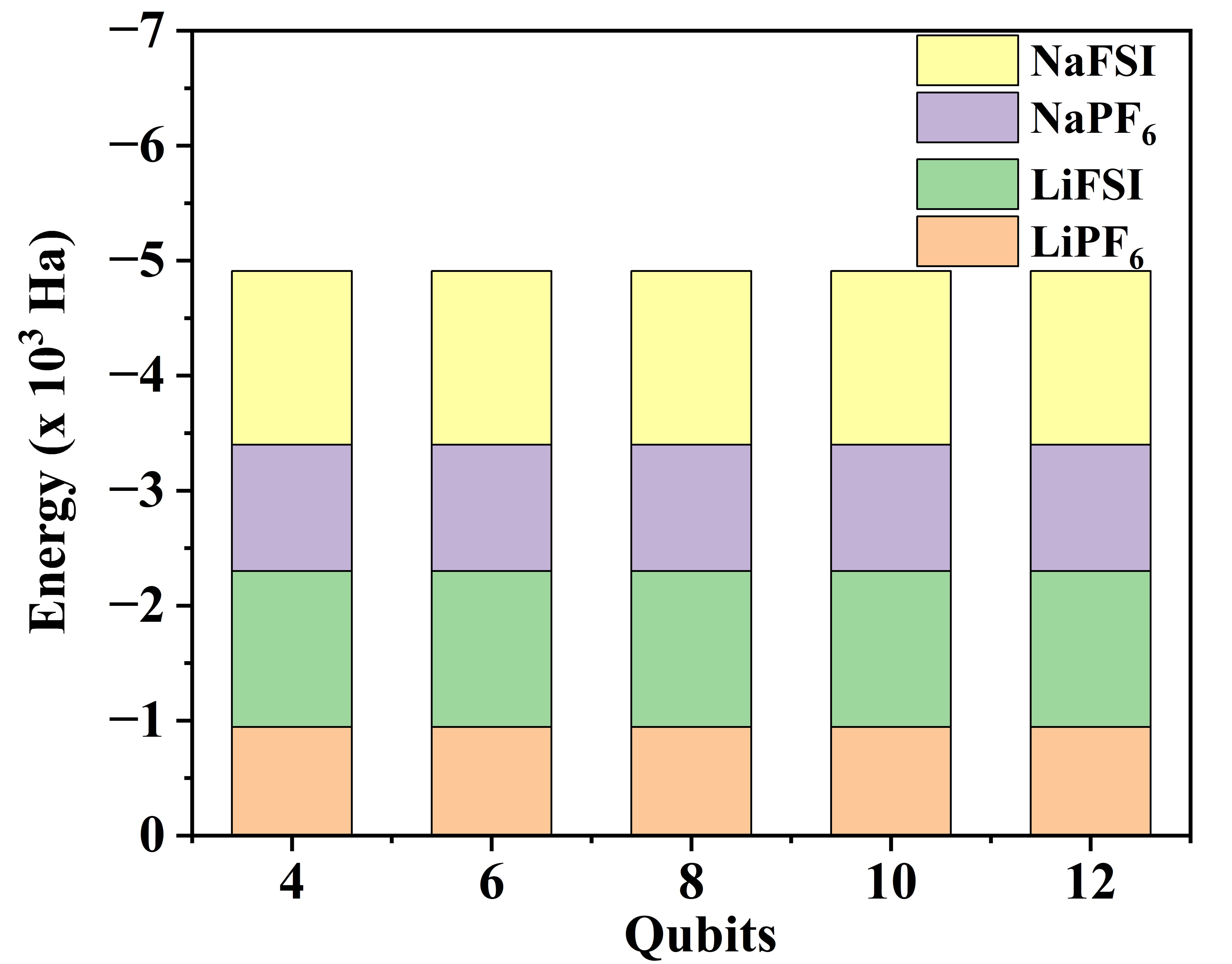}
    \caption{Energy vs Qubits for each molecules}
    \label{fig:qubit_reduction}
\end{figure}

Until recently, quantum computing for large molecular systems in chemistry has remained challenging and resource-intensive, primarily due to the limited qubit capacity of current hardware. Most studies have focused on small molecules such as H$_2$, LiH\cite{zhang2022variational}, BeH$_2$, and N$_2$, which typically require only 4 to 6 qubits\cite{ghosh2023deep}. Only a few investigations have ventured into more complex systems like Li$_2$S\cite{rice2021quantum}, phenylsulfonyl-carbazole TADF emitters\cite{gao2021applications}, C$_2$H$_4$\cite{cao2021towards}. For larger molecules, efficient qubit reduction strategies are essential to make quantum simulations practical. Figure~\ref{fig:qubit_reduction} presents the qubit optimization results for commercially relevant electrolyte salts: LiPF$_6$, NaPF$_6$, LiFSI, and NaFSI. Using the cc-pVDZ basis set, the unoptimized qubit requirements for these systems are prohibitively high, as shown in Table~\ref{tab:qubit_reduction}. However, through active space selection, we demonstrate that as few as 4 to 12 qubits can reproduce the same ground-state energy with negligible loss in accuracy. In our simulations, we selected 10 qubits corresponding to 6 valence electrons distributed over 5 orbitals (HOMO $\pm$ 2, see SI Figure S1), balancing accuracy and qubit efficiency.

\begin{table}[htbp]
\centering
\caption{Qubit requirements before and after tapering for each molecule using the cc-pVDZ basis set. The optimized qubit count corresponds to the smallest register size reproducing the full system's ground state energy.}
\label{tab:qubit_reduction}
\begin{tabular}{lcc}
\hline
\textbf{Molecule} & \textbf{Original Qubits} & \textbf{Optimized Qubits}$^{\dagger}$ \\
\hline
LiPF$_6$   & 232 & 10 \\
NaPF$_6$   & 240 & 10 \\
LiFSI       & 296 & 10 \\
NaFSI       & 304 & 10 \\
\hline
\end{tabular}

\vspace{1mm}
\raggedright
\footnotesize
$^{\dagger}$Qubits 4, 6, 8, 10, and 12 all yielded the same ground state energy (Fig. \ref{fig:qubit_reduction}); minimum qubit count shown.
\end{table}

\subsection{Basis Set Convergence Analysis}

\begin{figure}[htp]
    \centering
    \includegraphics[width=15cm]{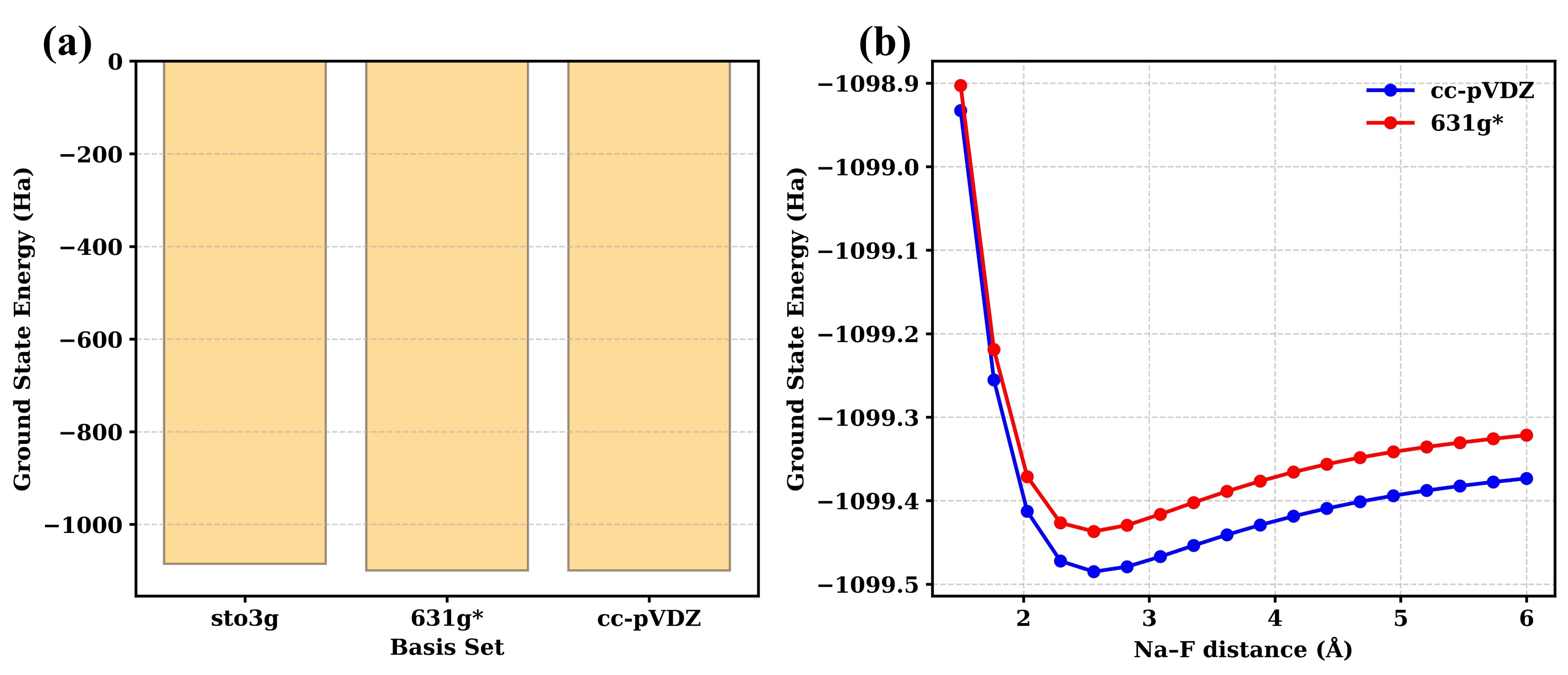}
    \caption{Basis set convergence for NaPF$_6$. (a) Ground-state energy using STO-3G, 6-31G*, and cc-pVDZ basis sets. (b) Relative energy profile along Na–F bond dissociation. 6-31G* and cc-pVDZ show good agreement, highlighting their reliability.}
    \label{fig:basis_convergence}
\end{figure}

Accurate ground-state and excited-state quantum simulations require an appropriate choice of basis set, as it significantly influences both the electronic structure and the computational cost. To evaluate the convergence of the basis set, we performed comparative simulations using three widely adopted basis sets STO-3G, 6-31G*, and cc-pVDZ. Figure~\ref{fig:basis_convergence} illustrates the convergence behavior for the NaPF$_6$ molecule, focusing on two key metrics: absolute ground-state energy (Figure \ref{fig:basis_convergence}a) and relative energy along the Na–F bond dissociation curve (Figure \ref{fig:basis_convergence}b).

In Figure \ref{fig:basis_convergence}(a), we observe a clear improvement in ground-state energy upon moving from the minimal STO-3G basis to the more flexible 6-31G* and cc-pVDZ basis sets, with the latter two yielding nearly identical energies. This indicates that both 6-31G* and cc-pVDZ sufficiently capture the electron correlation effects for this system. Figure \ref{fig:basis_convergence}(b) shows the relative potential energy profile as a function of Na–F bond length. Again, the curves for 6-31G* and cc-pVDZ closely overlap, confirming consistent energy predictions across bond dissociation coordinates, which is essential for simulating molecular stability and reactivity. Based on these results, we selected the cc-pVDZ basis set for all subsequent simulations in this study, as it provides a reliable balance between computational accuracy and cost. For the remaining electrolyte salts—LiPF$_6$, LiFSI, and NaFSI—similar basis set convergence tests were conducted and are included in the Supporting Information.

\subsection{Ansatz Benchmarking and Circuit Depth}

\begin{figure}[htp]
    \centering
    \includegraphics[width=16cm]{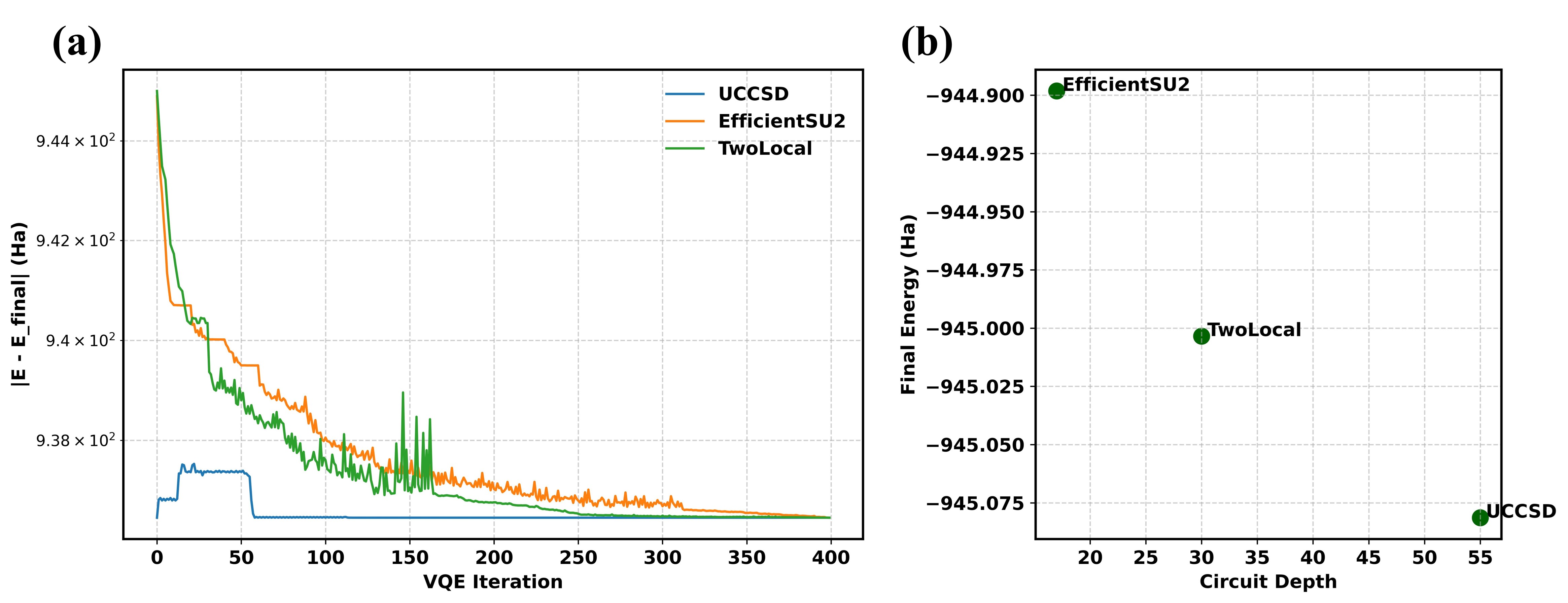}
    \caption{Comparison of VQE performance using \textit{UCCSD}, \textit{EfficientSU2}, and \textit{TwoLocal} ansätze for the LiPF$_6$ molecule. (a) Energy convergence with iterations. (b) Final energy vs. circuit depth, highlighting the trade-off between accuracy and resource requirements.}
    \label{fig:ansatz_comparison}
\end{figure}

To assess the efficiency and accuracy of different variational forms, we benchmarked three commonly used ansätze: UCCSD, EfficientSU2 and TwoLocal—on the LiPF$_6$ molecule using Qiskit's noiseless simulator. Figure~\ref{fig:ansatz_comparison}(a) shows the VQE convergence, while (b) compares final energies versus circuit depth and the CASCI reference (Exact (CASCI) = −945.0931 Ha, Table \ref{tab:vqe-comparison}). UCCSD achieves the lowest energy (−945.0814 Ha), closely matching Exact (CASCI), but requires the highest circuit depth (55), longest runtime (258.87 s), and 54 parameters. TwoLocal offers a balance with moderate accuracy (−945.0011 Ha), depth (30), and runtime (~21 s). EfficientSU2 converges fastest with minimal depth (17), but shows the highest energy deviation. These results underscore the critical trade-offs between accuracy, circuit depth, and computational overhead in VQE applications. While UCCSD remains ideal for simulations requiring high chemical precision, ansätze like TwoLocal may be more suitable for early quantum hardware deployments. It is important to note that all simulations were performed under noiseless conditions, and performance on real quantum devices is expected to degrade due to factors such as gate noise, readout errors, and decoherence. Therefore, the practical utility of each ansatz must be reevaluated in the context of realistic quantum noise and hardware limitations. While UCCSD delivers the best accuracy in our noiseless tests, shallower ansätze (TwoLocal/EfficientSU2) may be preferable on hardware due to reduced depth; in practice, we recommend pairing them with error mitigation and Pauli grouping to stabilize convergence.

\begin{table}[htbp]
\centering
\caption{Comparison of different variational ansätze used in VQE simulations for LiPF$_6$ molecule.}
\label{tab:vqe-comparison}
\begin{tabular}{lccccc}
\hline
\textbf{Ansatz} & \textbf{Qubits} & \textbf{Final Energy (Ha)} & \textbf{Depth} & \textbf{Parameters} & \textbf{Runtime (s)} \\
\hline
UCCSD           & 10  & $-945.0814$  & 55  & 54   & 259 \\
TwoLocal        & 10  & $-945.0011$  & 30  & 30   & 21   \\
EfficientSU2    & 10  & $-944.9738$  & 17  & 28   & 10   \\
\hline
CASCI (Reference) & 10  & $-945.0931$  & --  & --   & --  \\
\hline
\end{tabular}
\end{table}

\subsection{Dissociation Curves and Quantum Simulation Accuracy}

\begin{figure}[htp]
    \centering
    \includegraphics[width=15cm]{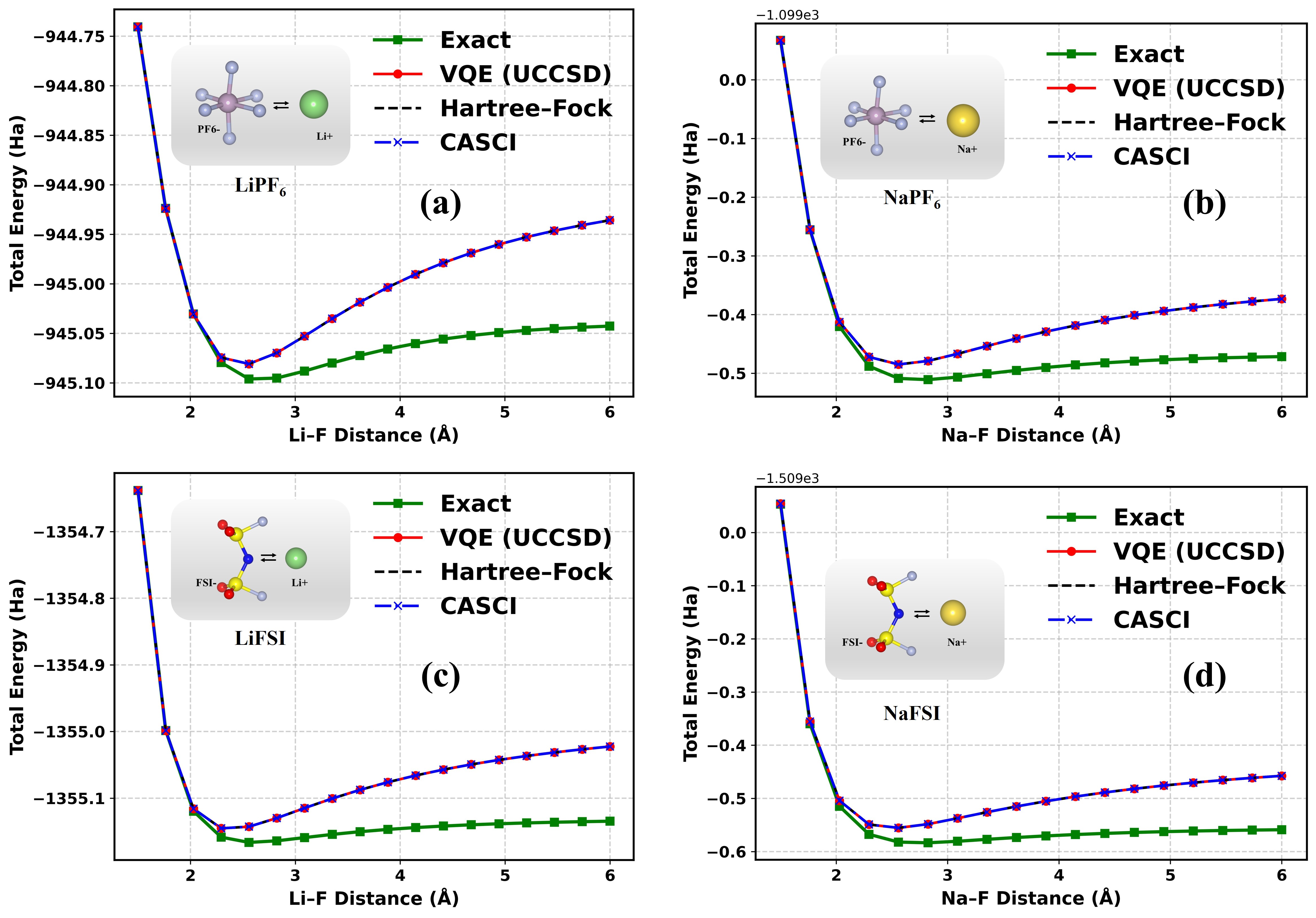}
    \caption{Dissociation energy curves of Li- and Na-based salts computed as a function of cation–anion distance. (a) LiPF$_6$, (b) NaPF$_6$, (c) LiFSI, and (d) NaFSI. The total energies were obtained using exact diagonalization (Exact), variational quantum eigensolver with UCCSD ansatz (VQE), Hartree–Fock (HF), and complete active space configuration interaction (CASCI). For both Li and Na salts, the equilibrium distances are shorter for Li–F bonds compared to Na–F, reflecting the smaller ionic radius of Li$^+$. FSI-based salts (LiFSI, NaFSI) exhibit deeper binding wells relative to PF$_6$-based salts, indicating stronger cation–anion interactions and higher thermodynamic stability.}
    \label{fig:dissociation_curve}
\end{figure}

The dissociation energy profiles of LiPF$_6$, LiFSI, NaPF$_6$, and NaFSI were computed as a function of the cation--anion distance to probe their binding characteristics and assess the accuracy of quantum simulations (Figure \ref{fig:dissociation_curve}). For all four electrolytes, the total energy decreases sharply as the alkali ion approaches the anion, reaching a minimum near $\sim$2.0~\AA{} for Li salts and $\sim$2.2~\AA{} for Na salts. These values correspond to the equilibrium Li--F and Na--F bond lengths, consistent with the smaller ionic radius of Li$^+$ relative to Na$^+$. Beyond these equilibrium distances, the energy gradually approaches a plateau, indicating complete cation--anion dissociation. Comparison of the energy minima shows that FSI-based salts (LiFSI, NaFSI) exhibit deeper wells than their PF$_6$ counterparts, reflecting stronger ionic binding and superior thermodynamic stability. Between the two cations, Li$^+$ generally forms stronger interactions with the anions than Na$^+$, in line with its higher charge density. Collectively, these results indicate that PF$_6$-based salts dissociate more readily, supporting higher ionic conductivity but reduced stability, while FSI-based salts bind more strongly, favoring stability at the expense of facile dissociation.

\begin{figure}[htp]
    \centering
    \includegraphics[width=15cm]{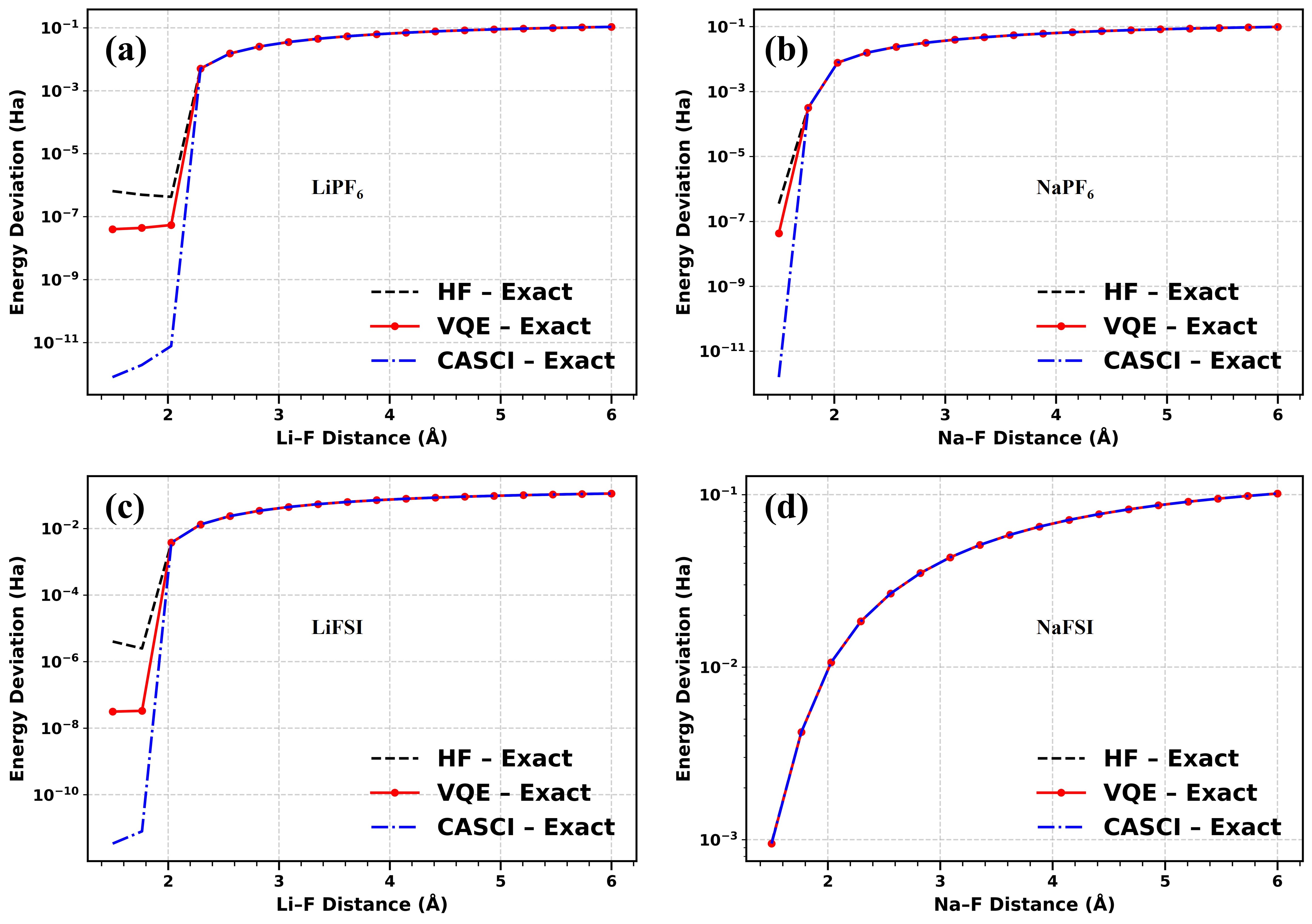}
    \caption{Energy deviation analysis of Li- and Na-based salts relative to exact diagonalization: (a) LiPF$_6$, (b) NaPF$_6$, (c) LiFSI, and (d) NaFSI. Deviations are shown for Hartree–Fock (HF), variational quantum eigensolver (VQE) with UCCSD ansatz, and complete active space configuration interaction (CASCI). VQE and CASCI closely follow the exact results with errors below $10^{-3}$ Ha, while HF shows larger deviations near equilibrium. Error growth at larger separations reflects increasing multi-reference character, underscoring the importance of quantum correlation methods for bond dissociation.}

    \label{fig:energy_deviations}
\end{figure}

To benchmark the precision of quantum algorithms, we evaluated the deviation of the approximate methods (Hartree--Fock, VQE with UCCSD and CASCI) from the exact reference energies throughout the complete bond-stretching regime with (Figure \ref{fig:energy_deviations}). The energy deviation plots reveal that VQE and CASCI closely reproduce the exact energies, achieving errors in the range of $10^{-7}$--$10^{-9}$~Ha near equilibrium bond distances. These error levels are well within the ``chemical accuracy'' threshold (1~kcal/mol $\approx 1.6 \times 10^{-3}$~Ha), demonstrating that quantum simulations can achieve predictive accuracy for electrolyte systems. In contrast, Hartree--Fock exhibits significantly larger deviations, particularly in the equilibrium region, underscoring its limitations in capturing electron correlation effects. As the distance between the cation and the anion increases, all methods show a gradual increase in error; however, VQE and CASCI remain below $\sim 10^{-2}$~Ha, confirming their robustness even in the strongly correlated dissociation limit.

The dissociation curve analysis and error benchmarks establish two key insights: (i) LiFSI and NaFSI are more stable salts than LiPF$_6$ and NaPF$_6$ due to their stronger cation--anion binding, and (ii) quantum computing approaches such as VQE with UCCSD are capable of reproducing these trends with near-exact accuracy. These findings highlight the dual importance of understanding electrolyte binding energetics and quantifying computational error limits, both of which are critical for guiding the design of next-generation lithium- and sodium-ion battery electrolytes using quantum simulations.

\subsection{Ground-State Benchmarks: VQE vs SQD}

To further assess the accuracy and scalability of our quantum simulations, we benchmarked the ground-state (GS) energies of LiPF$_6$, NaPF$_6$, LiFSI, and NaFSI salts using three different approaches: (i) exact diagonalization within the chosen active space (CASCI), (ii) VQE with a UCCSD ansatz on a reduced 10-qubit active space, and (iii) sample-based quantum diagonalization (SQD) with an extended 32-qubit active space (Table~\ref{tab:gs_energy}).

For all systems, the VQE (10-qubit) results reproduce the CASCI reference energies with deviations on the order of $10^{-4}$ Ha (sub-milliHartree), demonstrating that even a reduced active space captures the dominant electronic correlations relevant to the ground state. In particular, for LiPF$_6$, NaPF$_6$, and NaFSI, the deviations are below 0.05 mHa, which is well within chemical accuracy. A slightly larger discrepancy is observed for LiFSI ($\sim$2.4 mHa), reflecting the limitations of reduced active-space selection for systems with stronger correlation effects.

By contrast, SQD with a 32-qubit representation achieves near-perfect agreement with the exact CASCI results for all four salts, with differences consistently below $\sim$1.3 mHa and as small as $\sim$0.01 mHa in the case of NaPF$_6$. This highlights the systematic improvement gained by enlarging the active space, which allows SQD to recover essentially all correlation energy and serve as a high-accuracy benchmark.

These results illustrate the trade-off between efficiency and accuracy: while VQE with limited qubits provides chemically accurate results for most systems, SQD with larger qubit resources ensures near-exact benchmarking across diverse chemistries. Taken together with the dissociation curve analysis, this comparison underscores that both active space selection and qubit scaling critically determine quantum simulation performance, and that future quantum algorithms must balance hardware constraints with the need for chemically reliable accuracy.

\begin{table}[htbp]
\centering
\caption{Ground-state energies (in Hartree) of LiPF$_6$, NaPF$_6$, LiFSI, and NaFSI salts computed using exact diagonalization (CASCI), VQE with a 10-qubit active space, and SQD with a 32-qubit active space.}
\label{tab:gs_energy}
\begin{tabular}{lccccc}
\hline
\textbf{Method} & \textbf{LiPF$_6$} & \textbf{LiFSI} & \textbf{NaPF$_6$} & \textbf{NaFSI} \\
\hline
Exact (CASCI) & -945.081427 & -1355.148834 & -1099.485015 & -1509.555780 \\
VQE (10 qubit, 6e, 5o) & -945.081387 & -1355.146390 & -1099.484999 & -1509.555740 \\
SQD (32 qubit, 22e, 16o) & -945.081406 & -1355.147607 & -1099.485004 & -1509.555756 \\
\hline
\end{tabular}
\end{table}

With the conventional chemical accuracy target of approximately 1.6 mHa, SQD achieves chemical accuracy across all four systems. VQE with a 10-qubit active space now meets the chemical precision for LiPF$_6$, NaPF$_6$, and NaFSI; it remains marginal for LiFSI, where the limited active space introduces a noticeable deviation.

\begin{table}[h!]
\centering
\caption{Absolute deviations of ground-state energies (in mHa) from the CASCI reference for each molecule.}
\label{tab:error_mHa}
\begin{tabular}{lcc}
\hline
\textbf{Molecule} & \textbf{VQE (10 qubit)} & \textbf{SQD (32 qubit)} \\
\hline
LiPF$_6$  & 0.040  & 0.021 \\
LiFSI     & 2.444  & 1.227 \\
NaPF$_6$  & 0.016 & 0.011 \\
NaFSI     & 0.040  & 0.024 \\
\hline
\end{tabular}
\end{table}
SQD attains $\leq{1.3}$ mHa across all systems, whereas 10-qubit VQE is $\leq{0.05}$ mHa for LiPF$_6$, NaPF$_6$, NaFSI and 2.44 mHa for LiFSI, indicating that enlarging the active space systematically recovers correlation energy


\subsection{Excited-State Energies from VQE–qEOM }
The low-lying excited states of LiPF$_6$, NaPF$_6$, LiFSI, and NaFSI were computed using VQE combined with the quantum equation-of-motion (qEOM) approach. Table~\ref{tab:qEOM_main} summarizes the ground state and the first three excited states for each salt. For the PF$_6$-based electrolytes, the first excitations appear at $\sim$13.2 eV (LiPF$_6$) and $\sim$12.4 eV (NaPF$_6$), forming nearly degenerate manifolds. In contrast, the FSI-based salts exhibit lower excitation energies, with LiFSI and NaFSI showing first excitations near 8.8 eV and 8.4 eV, respectively. The splitting of these states into closely spaced pairs reflects the symmetry of the active spaces.
These results indicate that FSI-based salts are optically more accessible compared to PF$_6$ analogues, which is consistent with their distinct electronic environments and enhanced oxidative stability\cite{nishi2008electronic,wróbel2019interactions,berhaut2019ionic,wang2022liquid}. The ability of VQE–qEOM to reproduce chemically relevant excitation patterns demonstrates the promise of quantum algorithms for probing electrolyte photostability.

\begin{table*}[h!]
\centering
\caption{Ground-state energies and the first three excitation energies of LiPF$_6$, NaPF$_6$, LiFSI, and NaFSI obtained using VQE–qEOM.}
\label{tab:qEOM_main}
\resizebox{\textwidth}{!}{%
\begin{tabular}{lcccc}
\hline
\textbf{State} & \textbf{LiPF$_6$} & \textbf{NaPF$_6$} & \textbf{LiFSI} & \textbf{NaFSI} \\
 & Energy (Ha) / $\Delta E$ (eV) & Energy (Ha) / $\Delta E$ (eV) & Energy (Ha) / $\Delta E$ (eV) & Energy (Ha) / $\Delta E$ (eV) \\
\hline
Ground & -945.081387 / -- & -1099.485000 / -- & -1355.146390 / -- & -1509.555741 / -- \\
Excited 1 & -944.596961 / 13.18 & -1099.029210 / 12.40 & -1354.823457 / 8.79 & -1509.248406 / 8.36 \\
Excited 2 & -944.596757 / 13.19 & -1099.029004 / 12.41 & -1354.823056 / 8.80 & -1509.248116 / 8.37 \\
Excited 3 & -944.596643 / 13.19 & -1099.028848 / 12.41 & -1354.773106 / 10.16 & -1509.202515 / 9.61 \\
\hline
\end{tabular}}
\end{table*}

\begin{figure}[htp]
  \centering
  \includegraphics[width=15cm]{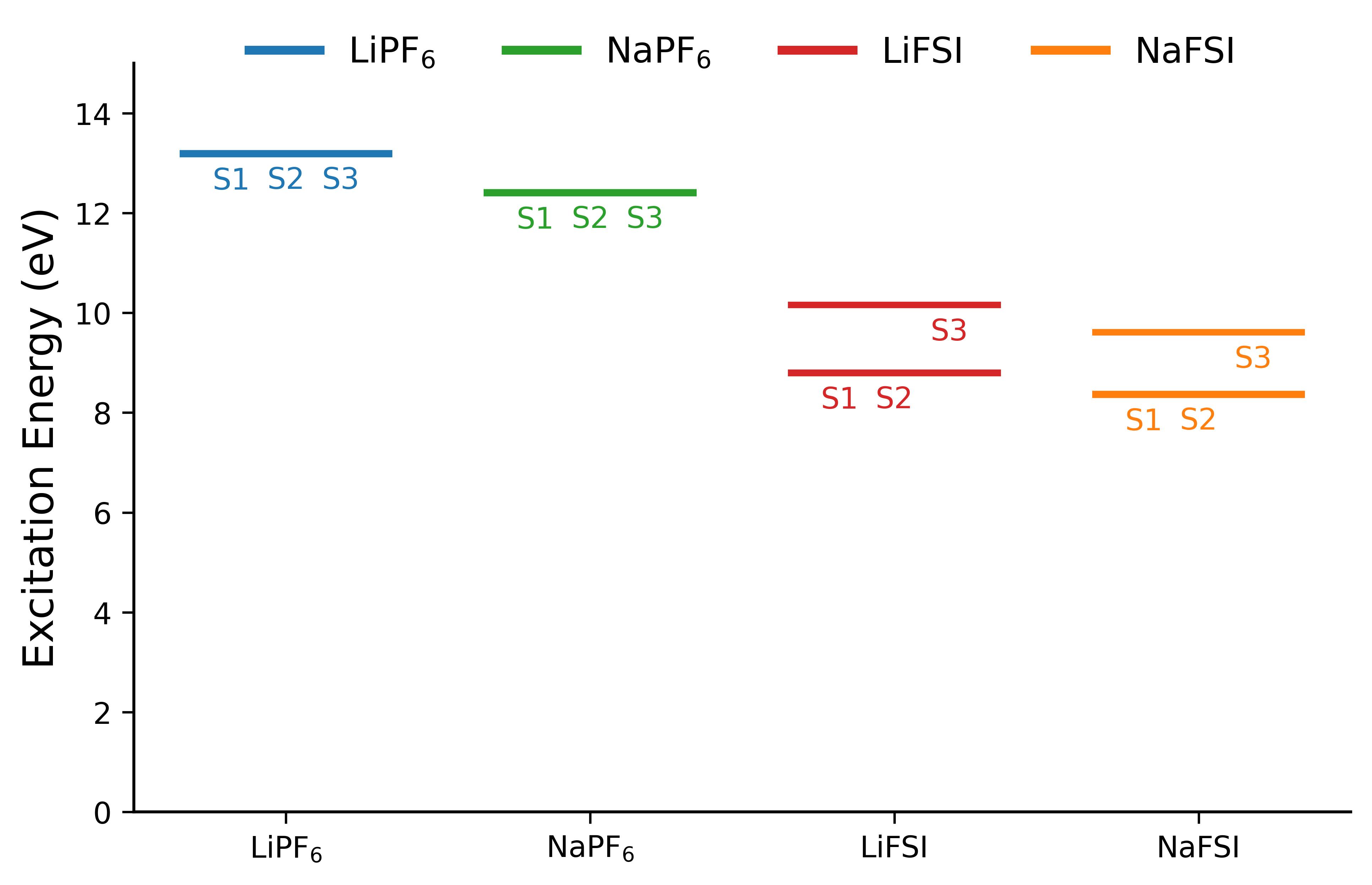}
  \caption{Lowest singlet excitation energies (S$_1$--S$_3$) from VQE–qEOM for LiPF$_6$, NaPF$_6$, LiFSI, and NaFSI. 
  FSI salts show markedly lower S$_1$ (narrower gaps) than PF$_6$. 
  PF$_6$ exhibits a near–degenerate cluster at 12–13~eV, whereas FSI presents a near–degenerate pair (S$_1$, S$_2$) and a higher S$_3$ separated by $\sim$1.3~eV. 
  Deep–UV onsets (nm): LiPF$_6$~$\sim$94, NaPF$_6$~$\sim$100, LiFSI~$\sim$141, NaFSI~$\sim$148.}
  \label{fig:exciton_energy}
\end{figure}

The VQE–qEOM spectra for LiPF$_6$, NaPF$_6$, LiFSI, and NaFSI reveal clear, chemically interpretable systematics. 
\emph{Anion effect:} FSI salts possess substantially lower first–excitation energies than their PF$_6$ analogues (Li: $13.18 \to 8.79$~eV, $\Delta\!\approx\!4.39$~eV; Na: $12.40 \to 8.36$~eV, $\Delta\!\approx\!4.04$~eV), implying narrower optical gaps for FSI. 
\emph{Cation effect:} Within a fixed anion, replacing Li$^+$ by Na$^+$ further decreases the first excitation (PF$_6$: $13.18 \to 12.40$~eV, $\Delta\!\approx\!0.78$~eV; FSI: $8.79 \to 8.36$~eV, $\Delta\!\approx\!0.43$~eV). 
\emph{Level structure:} PF$_6$ salts exhibit a compact three–state cluster at 12–13~eV (splittings $\sim$10~meV within rounding), whereas FSI salts display a near–degenerate pair (S$_1$, S$_2$) and a higher S$_3$ separated by $\sim$1.3~eV(Figure \ref{fig:exciton_energy}). 
Converting S$_1$ via $\lambda\,[\mathrm{nm}]=1240/E\,[\mathrm{eV}]$ places the absorption onsets in the deep UV (LiPF$_6$~$\approx$94~nm; NaPF$_6$~$\approx$100~nm; LiFSI~$\approx$141~nm; NaFSI~$\approx$148~nm), with FSI consistently red–shifted relative to PF$_6$. 
Taken together, these results indicate distinct anion electronic structures and a reproducible cation dependence.

The present analysis focuses on excitation energies; however, oscillator strengths are essential for assessing photostability and spectral intensity. 
Current qEOM implementations in Qiskit do not expose transition dipole moments directly. 
Future work will compute oscillator strengths from transition density matrices and analyze natural transition orbitals, enabling quantitative state assignments and brightness trends across the electrolyte series.

\section{Conclusion}
We have demonstrated that hybrid quantum algorithms can capture essential ground- and excited-state features of technologically relevant electrolyte salts. Using VQE for ground states and qEOM for vertical singlet excitations, supported by systematic active-space design, qubit reduction, and commuting-group measurements, we achieved close agreement with exact diagonalization within $\sim$10-qubit models. Extending to larger active spaces with sample-based quantum diagonalization (SQD) recovered near-exact (subspace-FCI) ground-state energies, providing a scalable benchmark.

Across LiPF$_6$, NaPF$_6$, LiFSI, and NaFSI, we identified robust spectral trends: (i) PF$_6$ salts possess higher first-excitation energies and a compact three-state manifold near 12--13~eV, whereas FSI salts exhibit lower onsets ($\sim$8--9~eV) with a (S$_1$,S$_2$) doublet and an S$_3$ $\sim$1.3~eV above; (ii) replacing Li$^+$ by Na$^+$ systematically narrows the optical gap by $\sim$0.4--0.8~eV within a given anion family; and (iii) dissociation profiles reflect stronger binding for FSI relative to PF$_6$ and for Li$^+$ relative to Na$^+$, consistent with ionic size and anion electronic structure. Taken together, these observations underscore distinct anion electronic environments and a reproducible cation dependence that are relevant to photostability and oxidative stability considerations.

The present study is limited to isolated species or embedded clusters compatible with NISQ resources; fully explicit solvation and interfacial effects remain out of reach for current hardware. In future work, we will integrate $\Delta$-solvation corrections (e.g., TDDFT/PCM or SMD), static or polarizable embedding into the one-electron operator, and transition-property evaluation (transition densities and natural transition orbitals) to obtain oscillator strengths and assign state characters. Hardware executions with calibrated error mitigation and expanded active spaces (including LUCJ/ADAPT ansätze and SKQD variants) will further test robustness. Overall, our results establish a reproducible quantum workflow for electrolyte screening and provide quantitative baselines to guide the rational design of next-generation battery materials.

\begin{acknowledgement}

This work was supported by the Korea Institute of Science and Technology (Grant number 2E31851), GKP (Global Knowledge Platform, Grant number 2V6760) project of the Ministry of Science, ICT and Future Planning. The authors thank Shubhayu Das of RWTH Aachen for useful discussions.

\end{acknowledgement}

\begin{suppinfo}

\end{suppinfo}

\paragraph{Data availability.}
All geometries, active-space definitions, qubit Hamiltonians, and scripts to reproduce the VQE-qEOM and SQD results will be deposited in a public repository upon acceptance.

\bibliography{achemso-demo}

\providecommand{\latin}[1]{#1}
\makeatletter
\providecommand{\doi}
  {\begingroup\let\do\@makeother\dospecials
  \catcode`\{=1 \catcode`\}=2 \doi@aux}
\providecommand{\doi@aux}[1]{\endgroup\texttt{#1}}
\makeatother
\providecommand*\mcitethebibliography{\thebibliography}
\csname @ifundefined\endcsname{endmcitethebibliography}  {\let\endmcitethebibliography\endthebibliography}{}
\begin{mcitethebibliography}{70}
\providecommand*\natexlab[1]{#1}
\providecommand*\mciteSetBstSublistMode[1]{}
\providecommand*\mciteSetBstMaxWidthForm[2]{}
\providecommand*\mciteBstWouldAddEndPuncttrue
  {\def\EndOfBibitem{\unskip.}}
\providecommand*\mciteBstWouldAddEndPunctfalse
  {\let\EndOfBibitem\relax}
\providecommand*\mciteSetBstMidEndSepPunct[3]{}
\providecommand*\mciteSetBstSublistLabelBeginEnd[3]{}
\providecommand*\EndOfBibitem{}
\mciteSetBstSublistMode{f}
\mciteSetBstMaxWidthForm{subitem}{(\alph{mcitesubitemcount})}
\mciteSetBstSublistLabelBeginEnd
  {\mcitemaxwidthsubitemform\space}
  {\relax}
  {\relax}

\bibitem[He \latin{et~al.}(2025)He, Ling, Wu, Lei, Cao, and Zhang]{he2025research}
He,~X.; Ling,~Y.; Wu,~Y.; Lei,~Y.; Cao,~D.; Zhang,~C. Research Progress of Electrolytes and Electrodes for Lithium-and Sodium-Ion Batteries at Extreme Temperatures. \emph{Small} \textbf{2025}, \emph{21}, 2412817\relax
\mciteBstWouldAddEndPuncttrue
\mciteSetBstMidEndSepPunct{\mcitedefaultmidpunct}
{\mcitedefaultendpunct}{\mcitedefaultseppunct}\relax
\EndOfBibitem
\bibitem[Patel \latin{et~al.}(2025)Patel, Jayakumar, Yen, and Izmaylov]{patel2025quantum}
Patel,~S.; Jayakumar,~P.; Yen,~T.-C.; Izmaylov,~A.~F. Quantum measurement for quantum chemistry on a quantum computer. \emph{Chemical Reviews} \textbf{2025}, \relax
\mciteBstWouldAddEndPunctfalse
\mciteSetBstMidEndSepPunct{\mcitedefaultmidpunct}
{}{\mcitedefaultseppunct}\relax
\EndOfBibitem
\bibitem[Cao \latin{et~al.}(2019)Cao, Romero, Olson, Degroote, Johnson, Kieferov{\'a}, Kivlichan, Menke, Peropadre, Sawaya, \latin{et~al.} others]{cao2019quantum}
Cao,~Y.; Romero,~J.; Olson,~J.~P.; Degroote,~M.; Johnson,~P.~D.; Kieferov{\'a},~M.; Kivlichan,~I.~D.; Menke,~T.; Peropadre,~B.; Sawaya,~N.~P.; others Quantum chemistry in the age of quantum computing. \emph{Chemical reviews} \textbf{2019}, \emph{119}, 10856--10915\relax
\mciteBstWouldAddEndPuncttrue
\mciteSetBstMidEndSepPunct{\mcitedefaultmidpunct}
{\mcitedefaultendpunct}{\mcitedefaultseppunct}\relax
\EndOfBibitem
\bibitem[Peruzzo \latin{et~al.}(2014)Peruzzo, McClean, Shadbolt, Yung, Zhou, Love, Aspuru-Guzik, and O’brien]{peruzzo2014variational}
Peruzzo,~A.; McClean,~J.; Shadbolt,~P.; Yung,~M.-H.; Zhou,~X.-Q.; Love,~P.~J.; Aspuru-Guzik,~A.; O’brien,~J.~L. A variational eigenvalue solver on a photonic quantum processor. \emph{Nature communications} \textbf{2014}, \emph{5}, 4213\relax
\mciteBstWouldAddEndPuncttrue
\mciteSetBstMidEndSepPunct{\mcitedefaultmidpunct}
{\mcitedefaultendpunct}{\mcitedefaultseppunct}\relax
\EndOfBibitem
\bibitem[Boyn \latin{et~al.}(2021)Boyn, Lykhin, Smart, Gagliardi, and Mazziotti]{boyn2021quantum}
Boyn,~J.-N.; Lykhin,~A.~O.; Smart,~S.~E.; Gagliardi,~L.; Mazziotti,~D.~A. Quantum-classical hybrid algorithm for the simulation of all-electron correlation. \emph{The Journal of Chemical Physics} \textbf{2021}, \emph{155}\relax
\mciteBstWouldAddEndPuncttrue
\mciteSetBstMidEndSepPunct{\mcitedefaultmidpunct}
{\mcitedefaultendpunct}{\mcitedefaultseppunct}\relax
\EndOfBibitem
\bibitem[McClean \latin{et~al.}(2016)McClean, Romero, Babbush, and Aspuru-Guzik]{mcclean2016theory}
McClean,~J.~R.; Romero,~J.; Babbush,~R.; Aspuru-Guzik,~A. The theory of variational hybrid quantum-classical algorithms. \emph{New Journal of Physics} \textbf{2016}, \emph{18}, 023023\relax
\mciteBstWouldAddEndPuncttrue
\mciteSetBstMidEndSepPunct{\mcitedefaultmidpunct}
{\mcitedefaultendpunct}{\mcitedefaultseppunct}\relax
\EndOfBibitem
\bibitem[Zheng \latin{et~al.}(2024)Zheng, Sun, Liu, Fan, Li, and Yang]{zheng2024quantum}
Zheng,~Y.; Sun,~Z.; Liu,~J.; Fan,~Y.; Li,~Z.; Yang,~J. Quantum Equation-of-Motion Method with Single, Double, and Triple Excitations. \emph{Journal of Chemical Theory and Computation} \textbf{2024}, \emph{20}, 9032--9040\relax
\mciteBstWouldAddEndPuncttrue
\mciteSetBstMidEndSepPunct{\mcitedefaultmidpunct}
{\mcitedefaultendpunct}{\mcitedefaultseppunct}\relax
\EndOfBibitem
\bibitem[Hlatshwayo \latin{et~al.}(2024)Hlatshwayo, Novak, and Litvinova]{hlatshwayo2024quantum}
Hlatshwayo,~M.~Q.; Novak,~J.; Litvinova,~E. Quantum benefit of the quantum equation of motion for the strongly coupled many-body problem. \emph{Physical Review C} \textbf{2024}, \emph{109}, 014306\relax
\mciteBstWouldAddEndPuncttrue
\mciteSetBstMidEndSepPunct{\mcitedefaultmidpunct}
{\mcitedefaultendpunct}{\mcitedefaultseppunct}\relax
\EndOfBibitem
\bibitem[Ollitrault \latin{et~al.}(2020)Ollitrault, Kandala, Chen, Barkoutsos, Mezzacapo, Pistoia, Sheldon, Woerner, Gambetta, and Tavernelli]{ollitrault2020quantum}
Ollitrault,~P.~J.; Kandala,~A.; Chen,~C.-F.; Barkoutsos,~P.~K.; Mezzacapo,~A.; Pistoia,~M.; Sheldon,~S.; Woerner,~S.; Gambetta,~J.~M.; Tavernelli,~I. Quantum equation of motion for computing molecular excitation energies on a noisy quantum processor. \emph{Physical Review Research} \textbf{2020}, \emph{2}, 043140\relax
\mciteBstWouldAddEndPuncttrue
\mciteSetBstMidEndSepPunct{\mcitedefaultmidpunct}
{\mcitedefaultendpunct}{\mcitedefaultseppunct}\relax
\EndOfBibitem
\bibitem[de~Gracia~Trivi{\~n}o \latin{et~al.}(2023)de~Gracia~Trivi{\~n}o, Delcey, and Wendin]{de2023complete}
de~Gracia~Trivi{\~n}o,~J.~A.; Delcey,~M.~G.; Wendin,~G. Complete active space methods for NISQ devices: The importance of canonical orbital optimization for accuracy and noise resilience. \emph{Journal of Chemical Theory and Computation} \textbf{2023}, \emph{19}, 2863--2872\relax
\mciteBstWouldAddEndPuncttrue
\mciteSetBstMidEndSepPunct{\mcitedefaultmidpunct}
{\mcitedefaultendpunct}{\mcitedefaultseppunct}\relax
\EndOfBibitem
\bibitem[Bharti \latin{et~al.}(2022)Bharti, Cervera-Lierta, Kyaw, Haug, Alperin-Lea, Anand, Degroote, Heimonen, Kottmann, Menke, \latin{et~al.} others]{bharti2022noisy}
Bharti,~K.; Cervera-Lierta,~A.; Kyaw,~T.~H.; Haug,~T.; Alperin-Lea,~S.; Anand,~A.; Degroote,~M.; Heimonen,~H.; Kottmann,~J.~S.; Menke,~T.; others Noisy intermediate-scale quantum algorithms. \emph{Reviews of Modern Physics} \textbf{2022}, \emph{94}, 015004\relax
\mciteBstWouldAddEndPuncttrue
\mciteSetBstMidEndSepPunct{\mcitedefaultmidpunct}
{\mcitedefaultendpunct}{\mcitedefaultseppunct}\relax
\EndOfBibitem
\bibitem[Acampora \latin{et~al.}(2023)Acampora, Di~Martino, Massa, Schiattarella, and Vitiello]{acampora2023d}
Acampora,~G.; Di~Martino,~F.; Massa,~A.; Schiattarella,~R.; Vitiello,~A. D-nisq: a reference model for distributed noisy intermediate-scale quantum computers. \emph{Information Fusion} \textbf{2023}, \emph{89}, 16--28\relax
\mciteBstWouldAddEndPuncttrue
\mciteSetBstMidEndSepPunct{\mcitedefaultmidpunct}
{\mcitedefaultendpunct}{\mcitedefaultseppunct}\relax
\EndOfBibitem
\bibitem[Robledo-Moreno \latin{et~al.}(2025)Robledo-Moreno, Motta, Haas, Javadi-Abhari, Jurcevic, Kirby, Martiel, Sharma, Sharma, Shirakawa, \latin{et~al.} others]{robledo2025chemistry}
Robledo-Moreno,~J.; Motta,~M.; Haas,~H.; Javadi-Abhari,~A.; Jurcevic,~P.; Kirby,~W.; Martiel,~S.; Sharma,~K.; Sharma,~S.; Shirakawa,~T.; others Chemistry beyond the scale of exact diagonalization on a quantum-centric supercomputer. \emph{Science Advances} \textbf{2025}, \emph{11}, eadu9991\relax
\mciteBstWouldAddEndPuncttrue
\mciteSetBstMidEndSepPunct{\mcitedefaultmidpunct}
{\mcitedefaultendpunct}{\mcitedefaultseppunct}\relax
\EndOfBibitem
\bibitem[Yu \latin{et~al.}(2025)Yu, Moreno, Iosue, Bertels, Claudino, Fuller, Groszkowski, Humble, Jurcevic, Kirby, \latin{et~al.} others]{yu2025quantum}
Yu,~J.; Moreno,~J.~R.; Iosue,~J.~T.; Bertels,~L.; Claudino,~D.; Fuller,~B.; Groszkowski,~P.; Humble,~T.~S.; Jurcevic,~P.; Kirby,~W.; others Quantum-centric algorithm for sample-based krylov diagonalization. \emph{arXiv preprint arXiv:2501.09702} \textbf{2025}, \relax
\mciteBstWouldAddEndPunctfalse
\mciteSetBstMidEndSepPunct{\mcitedefaultmidpunct}
{}{\mcitedefaultseppunct}\relax
\EndOfBibitem
\bibitem[Kanno \latin{et~al.}(2023)Kanno, Kohda, Imai, Koh, Mitarai, Mizukami, and Nakagawa]{kanno2023quantum}
Kanno,~K.; Kohda,~M.; Imai,~R.; Koh,~S.; Mitarai,~K.; Mizukami,~W.; Nakagawa,~Y.~O. Quantum-selected configuration interaction: Classical diagonalization of Hamiltonians in subspaces selected by quantum computers. \emph{arXiv preprint arXiv:2302.11320} \textbf{2023}, \relax
\mciteBstWouldAddEndPunctfalse
\mciteSetBstMidEndSepPunct{\mcitedefaultmidpunct}
{}{\mcitedefaultseppunct}\relax
\EndOfBibitem
\bibitem[Sugisaki \latin{et~al.}(2024)Sugisaki, Kanno, Itoko, Sakuma, and Yamamoto]{sugisaki2024hamiltonian}
Sugisaki,~K.; Kanno,~S.; Itoko,~T.; Sakuma,~R.; Yamamoto,~N. Hamiltonian simulation-based quantum-selected configuration interaction for large-scale electronic structure calculations with a quantum computer. \emph{arXiv preprint arXiv:2412.07218} \textbf{2024}, \relax
\mciteBstWouldAddEndPunctfalse
\mciteSetBstMidEndSepPunct{\mcitedefaultmidpunct}
{}{\mcitedefaultseppunct}\relax
\EndOfBibitem
\bibitem[Sureshbabu \latin{et~al.}(2021)Sureshbabu, Sajjan, Oh, and Kais]{sureshbabu2021implementation}
Sureshbabu,~S.~H.; Sajjan,~M.; Oh,~S.; Kais,~S. Implementation of quantum machine learning for electronic structure calculations of periodic systems on quantum computing devices. \emph{Journal of Chemical Information and Modeling} \textbf{2021}, \emph{61}, 2667--2674\relax
\mciteBstWouldAddEndPuncttrue
\mciteSetBstMidEndSepPunct{\mcitedefaultmidpunct}
{\mcitedefaultendpunct}{\mcitedefaultseppunct}\relax
\EndOfBibitem
\bibitem[O’Malley \latin{et~al.}(2016)O’Malley, Babbush, Kivlichan, Romero, McClean, Barends, Kelly, Roushan, Tranter, Ding, \latin{et~al.} others]{o2016scalable}
O’Malley,~P.~J.; Babbush,~R.; Kivlichan,~I.~D.; Romero,~J.; McClean,~J.~R.; Barends,~R.; Kelly,~J.; Roushan,~P.; Tranter,~A.; Ding,~N.; others Scalable quantum simulation of molecular energies. \emph{Physical Review X} \textbf{2016}, \emph{6}, 031007\relax
\mciteBstWouldAddEndPuncttrue
\mciteSetBstMidEndSepPunct{\mcitedefaultmidpunct}
{\mcitedefaultendpunct}{\mcitedefaultseppunct}\relax
\EndOfBibitem
\bibitem[Ghosh \latin{et~al.}(2023)Ghosh, Kumar, Rajan, and Yamijala]{ghosh2023deep}
Ghosh,~K.; Kumar,~S.; Rajan,~N.~M.; Yamijala,~S.~S. Deep neural network assisted quantum chemistry calculations on quantum computers. \emph{ACS omega} \textbf{2023}, \emph{8}, 48211--48220\relax
\mciteBstWouldAddEndPuncttrue
\mciteSetBstMidEndSepPunct{\mcitedefaultmidpunct}
{\mcitedefaultendpunct}{\mcitedefaultseppunct}\relax
\EndOfBibitem
\bibitem[Tao \latin{et~al.}(2022)Tao, Zeng, Fan, Liu, Li, and Yang]{tao2022exploring}
Tao,~Y.; Zeng,~X.; Fan,~Y.; Liu,~J.; Li,~Z.; Yang,~J. Exploring accurate potential energy surfaces via integrating Variational Quantum Eigensolver with machine learning. \emph{The Journal of Physical Chemistry Letters} \textbf{2022}, \emph{13}, 6420--6426\relax
\mciteBstWouldAddEndPuncttrue
\mciteSetBstMidEndSepPunct{\mcitedefaultmidpunct}
{\mcitedefaultendpunct}{\mcitedefaultseppunct}\relax
\EndOfBibitem
\bibitem[L{\"o}tstedt \latin{et~al.}(2022)L{\"o}tstedt, Yamanouchi, and Tachikawa]{lotstedt2022evaluation}
L{\"o}tstedt,~E.; Yamanouchi,~K.; Tachikawa,~Y. Evaluation of vibrational energies and wave functions of CO2 on a quantum computer. \emph{AVS Quantum Science} \textbf{2022}, \emph{4}\relax
\mciteBstWouldAddEndPuncttrue
\mciteSetBstMidEndSepPunct{\mcitedefaultmidpunct}
{\mcitedefaultendpunct}{\mcitedefaultseppunct}\relax
\EndOfBibitem
\bibitem[Bauer \latin{et~al.}(2020)Bauer, Bravyi, Motta, and Chan]{bauer2020quantum}
Bauer,~B.; Bravyi,~S.; Motta,~M.; Chan,~G. K.-L. Quantum algorithms for quantum chemistry and quantum materials science. \emph{Chemical reviews} \textbf{2020}, \emph{120}, 12685--12717\relax
\mciteBstWouldAddEndPuncttrue
\mciteSetBstMidEndSepPunct{\mcitedefaultmidpunct}
{\mcitedefaultendpunct}{\mcitedefaultseppunct}\relax
\EndOfBibitem
\bibitem[Blunt \latin{et~al.}(2022)Blunt, Camps, Crawford, Izs{\'a}k, Leontica, Mirani, Moylett, Scivier, Sunderhauf, Schopf, \latin{et~al.} others]{blunt2022perspective}
Blunt,~N.~S.; Camps,~J.; Crawford,~O.; Izs{\'a}k,~R.; Leontica,~S.; Mirani,~A.; Moylett,~A.~E.; Scivier,~S.~A.; Sunderhauf,~C.; Schopf,~P.; others Perspective on the current state-of-the-art of quantum computing for drug discovery applications. \emph{Journal of Chemical Theory and Computation} \textbf{2022}, \emph{18}, 7001--7023\relax
\mciteBstWouldAddEndPuncttrue
\mciteSetBstMidEndSepPunct{\mcitedefaultmidpunct}
{\mcitedefaultendpunct}{\mcitedefaultseppunct}\relax
\EndOfBibitem
\bibitem[Cava \latin{et~al.}(2021)Cava, De~Leon, and Xie]{cava2021introduction}
Cava,~R.; De~Leon,~N.; Xie,~W. Introduction: quantum materials. 2021\relax
\mciteBstWouldAddEndPuncttrue
\mciteSetBstMidEndSepPunct{\mcitedefaultmidpunct}
{\mcitedefaultendpunct}{\mcitedefaultseppunct}\relax
\EndOfBibitem
\bibitem[Delgado-Granados \latin{et~al.}(2025)Delgado-Granados, Krogmeier, Sager-Smith, Avdic, Hu, Sajjan, Abbasi, Smart, Narang, Kais, \latin{et~al.} others]{delgado2025quantum}
Delgado-Granados,~L.~H.; Krogmeier,~T.~J.; Sager-Smith,~L.~M.; Avdic,~I.; Hu,~Z.; Sajjan,~M.; Abbasi,~M.; Smart,~S.~E.; Narang,~P.; Kais,~S.; others Quantum algorithms and applications for open quantum systems. \emph{Chemical Reviews} \textbf{2025}, \emph{125}, 1823--1839\relax
\mciteBstWouldAddEndPuncttrue
\mciteSetBstMidEndSepPunct{\mcitedefaultmidpunct}
{\mcitedefaultendpunct}{\mcitedefaultseppunct}\relax
\EndOfBibitem
\bibitem[Hariharan \latin{et~al.}(2024)Hariharan, Kinge, and Visscher]{hariharan2024modeling}
Hariharan,~S.; Kinge,~S.; Visscher,~L. Modeling heterogeneous catalysis using quantum computers: An academic and industry perspective. \emph{Journal of chemical information and modeling} \textbf{2024}, \emph{65}, 472--511\relax
\mciteBstWouldAddEndPuncttrue
\mciteSetBstMidEndSepPunct{\mcitedefaultmidpunct}
{\mcitedefaultendpunct}{\mcitedefaultseppunct}\relax
\EndOfBibitem
\bibitem[Lipka-Bartosik \latin{et~al.}(2024)Lipka-Bartosik, Wilming, and Ng]{lipka2024catalysis}
Lipka-Bartosik,~P.; Wilming,~H.; Ng,~N.~H. Catalysis in quantum information theory. \emph{Reviews of Modern Physics} \textbf{2024}, \emph{96}, 025005\relax
\mciteBstWouldAddEndPuncttrue
\mciteSetBstMidEndSepPunct{\mcitedefaultmidpunct}
{\mcitedefaultendpunct}{\mcitedefaultseppunct}\relax
\EndOfBibitem
\bibitem[von Burg \latin{et~al.}(2021)von Burg, Low, H{\"a}ner, Steiger, Reiher, Roetteler, and Troyer]{von2021quantum}
von Burg,~V.; Low,~G.~H.; H{\"a}ner,~T.; Steiger,~D.~S.; Reiher,~M.; Roetteler,~M.; Troyer,~M. Quantum computing enhanced computational catalysis. \emph{Physical Review Research} \textbf{2021}, \emph{3}, 033055\relax
\mciteBstWouldAddEndPuncttrue
\mciteSetBstMidEndSepPunct{\mcitedefaultmidpunct}
{\mcitedefaultendpunct}{\mcitedefaultseppunct}\relax
\EndOfBibitem
\bibitem[Motta \latin{et~al.}(2020)Motta, Gujarati, Rice, Kumar, Masteran, Latone, Lee, Valeev, and Takeshita]{motta2020quantum}
Motta,~M.; Gujarati,~T.~P.; Rice,~J.~E.; Kumar,~A.; Masteran,~C.; Latone,~J.~A.; Lee,~E.; Valeev,~E.~F.; Takeshita,~T.~Y. Quantum simulation of electronic structure with a transcorrelated Hamiltonian: improved accuracy with a smaller footprint on the quantum computer. \emph{Physical Chemistry Chemical Physics} \textbf{2020}, \emph{22}, 24270--24281\relax
\mciteBstWouldAddEndPuncttrue
\mciteSetBstMidEndSepPunct{\mcitedefaultmidpunct}
{\mcitedefaultendpunct}{\mcitedefaultseppunct}\relax
\EndOfBibitem
\bibitem[Barkoutsos \latin{et~al.}(2018)Barkoutsos, Gonthier, Sokolov, Moll, Salis, Fuhrer, Ganzhorn, Egger, Troyer, Mezzacapo, \latin{et~al.} others]{barkoutsos2018quantum}
Barkoutsos,~P.~K.; Gonthier,~J.~F.; Sokolov,~I.; Moll,~N.; Salis,~G.; Fuhrer,~A.; Ganzhorn,~M.; Egger,~D.~J.; Troyer,~M.; Mezzacapo,~A.; others Quantum algorithms for electronic structure calculations: Particle-hole Hamiltonian and optimized wave-function expansions. \emph{Physical Review A} \textbf{2018}, \emph{98}, 022322\relax
\mciteBstWouldAddEndPuncttrue
\mciteSetBstMidEndSepPunct{\mcitedefaultmidpunct}
{\mcitedefaultendpunct}{\mcitedefaultseppunct}\relax
\EndOfBibitem
\bibitem[Ollitrault \latin{et~al.}(2020)Ollitrault, Mazzola, and Tavernelli]{ollitrault2020nonadiabatic}
Ollitrault,~P.~J.; Mazzola,~G.; Tavernelli,~I. Nonadiabatic molecular quantum dynamics with quantum computers. \emph{Physical Review Letters} \textbf{2020}, \emph{125}, 260511\relax
\mciteBstWouldAddEndPuncttrue
\mciteSetBstMidEndSepPunct{\mcitedefaultmidpunct}
{\mcitedefaultendpunct}{\mcitedefaultseppunct}\relax
\EndOfBibitem
\bibitem[Miessen \latin{et~al.}(2021)Miessen, Ollitrault, and Tavernelli]{miessen2021quantum}
Miessen,~A.; Ollitrault,~P.~J.; Tavernelli,~I. Quantum algorithms for quantum dynamics: A performance study on the spin-boson model. \emph{Physical Review Research} \textbf{2021}, \emph{3}, 043212\relax
\mciteBstWouldAddEndPuncttrue
\mciteSetBstMidEndSepPunct{\mcitedefaultmidpunct}
{\mcitedefaultendpunct}{\mcitedefaultseppunct}\relax
\EndOfBibitem
\bibitem[Paudel \latin{et~al.}(2022)Paudel, Syamlal, Crawford, Lee, Shugayev, Lu, Ohodnicki, Mollot, and Duan]{paudel2022quantum}
Paudel,~H.~P.; Syamlal,~M.; Crawford,~S.~E.; Lee,~Y.-L.; Shugayev,~R.~A.; Lu,~P.; Ohodnicki,~P.~R.; Mollot,~D.; Duan,~Y. Quantum computing and simulations for energy applications: Review and perspective. \emph{ACS Engineering Au} \textbf{2022}, \emph{2}, 151--196\relax
\mciteBstWouldAddEndPuncttrue
\mciteSetBstMidEndSepPunct{\mcitedefaultmidpunct}
{\mcitedefaultendpunct}{\mcitedefaultseppunct}\relax
\EndOfBibitem
\bibitem[McArdle \latin{et~al.}(2020)McArdle, Endo, Aspuru-Guzik, Benjamin, and Yuan]{mcardle2020quantum}
McArdle,~S.; Endo,~S.; Aspuru-Guzik,~A.; Benjamin,~S.~C.; Yuan,~X. Quantum computational chemistry. \emph{Reviews of Modern Physics} \textbf{2020}, \emph{92}, 015003\relax
\mciteBstWouldAddEndPuncttrue
\mciteSetBstMidEndSepPunct{\mcitedefaultmidpunct}
{\mcitedefaultendpunct}{\mcitedefaultseppunct}\relax
\EndOfBibitem
\bibitem[Rice \latin{et~al.}(2021)Rice, Gujarati, Motta, Takeshita, Lee, Latone, and Garcia]{rice2021quantum}
Rice,~J.~E.; Gujarati,~T.~P.; Motta,~M.; Takeshita,~T.~Y.; Lee,~E.; Latone,~J.~A.; Garcia,~J.~M. Quantum computation of dominant products in lithium--sulfur batteries. \emph{The Journal of Chemical Physics} \textbf{2021}, \emph{154}\relax
\mciteBstWouldAddEndPuncttrue
\mciteSetBstMidEndSepPunct{\mcitedefaultmidpunct}
{\mcitedefaultendpunct}{\mcitedefaultseppunct}\relax
\EndOfBibitem
\bibitem[Gao \latin{et~al.}(2021)Gao, Jones, Motta, Sugawara, Watanabe, Kobayashi, Watanabe, Ohnishi, Nakamura, and Yamamoto]{gao2021applications}
Gao,~Q.; Jones,~G.~O.; Motta,~M.; Sugawara,~M.; Watanabe,~H.~C.; Kobayashi,~T.; Watanabe,~E.; Ohnishi,~Y.-y.; Nakamura,~H.; Yamamoto,~N. Applications of quantum computing for investigations of electronic transitions in phenylsulfonyl-carbazole TADF emitters. \emph{npj Computational Materials} \textbf{2021}, \emph{7}, 70\relax
\mciteBstWouldAddEndPuncttrue
\mciteSetBstMidEndSepPunct{\mcitedefaultmidpunct}
{\mcitedefaultendpunct}{\mcitedefaultseppunct}\relax
\EndOfBibitem
\bibitem[Treinish(2023)]{matthew_treinish_2023_8190968}
Treinish,~M. Qiskit/qiskit-metapackage: Qiskit 0.44.0. 2023; \url{https://doi.org/10.5281/zenodo.8190968}\relax
\mciteBstWouldAddEndPuncttrue
\mciteSetBstMidEndSepPunct{\mcitedefaultmidpunct}
{\mcitedefaultendpunct}{\mcitedefaultseppunct}\relax
\EndOfBibitem
\bibitem[Frisch \latin{et~al.}(2016)Frisch, Trucks, Schlegel, Scuseria, Robb, Cheeseman, Scalmani, Barone, Petersson, Nakatsuji, Li, Caricato, Marenich, Bloino, Janesko, Gomperts, Mennucci, Hratchian, Ortiz, Izmaylov, Sonnenberg, Williams-Young, Ding, Lipparini, Egidi, Goings, Peng, Petrone, Henderson, Ranasinghe, Zakrzewski, Gao, Rega, Zheng, Liang, Hada, Ehara, Toyota, Fukuda, Hasegawa, Ishida, Nakajima, Honda, Kitao, Nakai, Vreven, Throssell, Montgomery, Peralta, Ogliaro, Bearpark, Heyd, Brothers, Kudin, Staroverov, Keith, Kobayashi, Normand, Raghavachari, Rendell, Burant, Iyengar, Tomasi, Cossi, Millam, Klene, Adamo, Cammi, Ochterski, Martin, Morokuma, Farkas, Foresman, and Fox]{g16}
Frisch,~M.~J. \latin{et~al.}  Gaussian˜16 {R}evision {C}.01. 2016; Gaussian Inc. Wallingford CT\relax
\mciteBstWouldAddEndPuncttrue
\mciteSetBstMidEndSepPunct{\mcitedefaultmidpunct}
{\mcitedefaultendpunct}{\mcitedefaultseppunct}\relax
\EndOfBibitem
\bibitem[Kaliakin \latin{et~al.}(2025)Kaliakin, Shajan, Liang, and Merz~Jr]{kaliakin2025implicit}
Kaliakin,~D.; Shajan,~A.; Liang,~F.; Merz~Jr,~K.~M. Implicit solvent sample-based quantum diagonalization. \emph{The Journal of Physical Chemistry B} \textbf{2025}, \emph{129}, 5788--5796\relax
\mciteBstWouldAddEndPuncttrue
\mciteSetBstMidEndSepPunct{\mcitedefaultmidpunct}
{\mcitedefaultendpunct}{\mcitedefaultseppunct}\relax
\EndOfBibitem
\bibitem[Tilly \latin{et~al.}(2022)Tilly, Chen, Cao, Picozzi, Setia, Li, Grant, Wossnig, Rungger, Booth, \latin{et~al.} others]{tilly2022variational}
Tilly,~J.; Chen,~H.; Cao,~S.; Picozzi,~D.; Setia,~K.; Li,~Y.; Grant,~E.; Wossnig,~L.; Rungger,~I.; Booth,~G.~H.; others The variational quantum eigensolver: a review of methods and best practices. \emph{Physics Reports} \textbf{2022}, \emph{986}, 1--128\relax
\mciteBstWouldAddEndPuncttrue
\mciteSetBstMidEndSepPunct{\mcitedefaultmidpunct}
{\mcitedefaultendpunct}{\mcitedefaultseppunct}\relax
\EndOfBibitem
\bibitem[Sun \latin{et~al.}(2020)Sun, Zhang, Banerjee, Bao, Barbry, Blunt, Bogdanov, Booth, Chen, Cui, \latin{et~al.} others]{sun2020recent}
Sun,~Q.; Zhang,~X.; Banerjee,~S.; Bao,~P.; Barbry,~M.; Blunt,~N.~S.; Bogdanov,~N.~A.; Booth,~G.~H.; Chen,~J.; Cui,~Z.-H.; others Recent developments in the PySCF program package. \emph{The Journal of chemical physics} \textbf{2020}, \emph{153}\relax
\mciteBstWouldAddEndPuncttrue
\mciteSetBstMidEndSepPunct{\mcitedefaultmidpunct}
{\mcitedefaultendpunct}{\mcitedefaultseppunct}\relax
\EndOfBibitem
\bibitem[Sun \latin{et~al.}(2018)Sun, Berkelbach, Blunt, Booth, Guo, Li, Liu, McClain, Sayfutyarova, Sharma, \latin{et~al.} others]{sun2018pyscf}
Sun,~Q.; Berkelbach,~T.~C.; Blunt,~N.~S.; Booth,~G.~H.; Guo,~S.; Li,~Z.; Liu,~J.; McClain,~J.~D.; Sayfutyarova,~E.~R.; Sharma,~S.; others PySCF: the Python-based simulations of chemistry framework. \emph{Wiley Interdisciplinary Reviews: Computational Molecular Science} \textbf{2018}, \emph{8}, e1340\relax
\mciteBstWouldAddEndPuncttrue
\mciteSetBstMidEndSepPunct{\mcitedefaultmidpunct}
{\mcitedefaultendpunct}{\mcitedefaultseppunct}\relax
\EndOfBibitem
\bibitem[Ekstrom \latin{et~al.}(2010)Ekstrom, Visscher, Bast, Thorvaldsen, and Ruud]{ekström2010arbitrary}
Ekstrom,~U.; Visscher,~L.; Bast,~R.; Thorvaldsen,~A.~J.; Ruud,~K. Arbitrary-order density functional response theory from automatic differentiation. \emph{Journal of chemical theory and computation} \textbf{2010}, \emph{6}, 1971--1980\relax
\mciteBstWouldAddEndPuncttrue
\mciteSetBstMidEndSepPunct{\mcitedefaultmidpunct}
{\mcitedefaultendpunct}{\mcitedefaultseppunct}\relax
\EndOfBibitem
\bibitem[Jordan and Wigner(1928)Jordan, and Wigner]{jordan1928pauli}
Jordan,~P.; Wigner,~E.~P. About the Pauli exclusion principle. \emph{Z. Phys} \textbf{1928}, \emph{47}, 14--75\relax
\mciteBstWouldAddEndPuncttrue
\mciteSetBstMidEndSepPunct{\mcitedefaultmidpunct}
{\mcitedefaultendpunct}{\mcitedefaultseppunct}\relax
\EndOfBibitem
\bibitem[Ortiz \latin{et~al.}(2002)Ortiz, Gubernatis, Knill, and Laflamme]{ortiz2002simulating}
Ortiz,~G.; Gubernatis,~J.; Knill,~E.; Laflamme,~R. Simulating fermions on a quantum computer. \emph{Computer physics communications} \textbf{2002}, \emph{146}, 302--316\relax
\mciteBstWouldAddEndPuncttrue
\mciteSetBstMidEndSepPunct{\mcitedefaultmidpunct}
{\mcitedefaultendpunct}{\mcitedefaultseppunct}\relax
\EndOfBibitem
\bibitem[Somma \latin{et~al.}(2002)Somma, Ortiz, Gubernatis, Knill, and Laflamme]{somma2002simulating}
Somma,~R.; Ortiz,~G.; Gubernatis,~J.~E.; Knill,~E.; Laflamme,~R. Simulating physical phenomena by quantum networks. \emph{Physical Review A} \textbf{2002}, \emph{65}, 042323\relax
\mciteBstWouldAddEndPuncttrue
\mciteSetBstMidEndSepPunct{\mcitedefaultmidpunct}
{\mcitedefaultendpunct}{\mcitedefaultseppunct}\relax
\EndOfBibitem
\bibitem[Grimsley \latin{et~al.}(2019)Grimsley, Economou, Barnes, and Mayhall]{grimsley2019adaptive}
Grimsley,~H.~R.; Economou,~S.~E.; Barnes,~E.; Mayhall,~N.~J. An adaptive variational algorithm for exact molecular simulations on a quantum computer. \emph{Nature communications} \textbf{2019}, \emph{10}, 3007\relax
\mciteBstWouldAddEndPuncttrue
\mciteSetBstMidEndSepPunct{\mcitedefaultmidpunct}
{\mcitedefaultendpunct}{\mcitedefaultseppunct}\relax
\EndOfBibitem
\bibitem[Romero \latin{et~al.}(2018)Romero, Babbush, McClean, Hempel, Love, and Aspuru-Guzik]{romero2018strategies}
Romero,~J.; Babbush,~R.; McClean,~J.~R.; Hempel,~C.; Love,~P.~J.; Aspuru-Guzik,~A. Strategies for quantum computing molecular energies using the unitary coupled cluster ansatz. \emph{Quantum Science and Technology} \textbf{2018}, \emph{4}, 014008\relax
\mciteBstWouldAddEndPuncttrue
\mciteSetBstMidEndSepPunct{\mcitedefaultmidpunct}
{\mcitedefaultendpunct}{\mcitedefaultseppunct}\relax
\EndOfBibitem
\bibitem[Grimsley \latin{et~al.}(2019)Grimsley, Claudino, Economou, Barnes, and Mayhall]{grimsley2019trotterized}
Grimsley,~H.~R.; Claudino,~D.; Economou,~S.~E.; Barnes,~E.; Mayhall,~N.~J. Is the trotterized uccsd ansatz chemically well-defined? \emph{Journal of chemical theory and computation} \textbf{2019}, \emph{16}, 1--6\relax
\mciteBstWouldAddEndPuncttrue
\mciteSetBstMidEndSepPunct{\mcitedefaultmidpunct}
{\mcitedefaultendpunct}{\mcitedefaultseppunct}\relax
\EndOfBibitem
\bibitem[Evangelista \latin{et~al.}(2019)Evangelista, Chan, and Scuseria]{evangelista2019exact}
Evangelista,~F.~A.; Chan,~G.~K.; Scuseria,~G.~E. Exact parameterization of fermionic wave functions via unitary coupled cluster theory. \emph{The Journal of chemical physics} \textbf{2019}, \emph{151}\relax
\mciteBstWouldAddEndPuncttrue
\mciteSetBstMidEndSepPunct{\mcitedefaultmidpunct}
{\mcitedefaultendpunct}{\mcitedefaultseppunct}\relax
\EndOfBibitem
\bibitem[Kandala \latin{et~al.}(2017)Kandala, Mezzacapo, Temme, Takita, Brink, Chow, and Gambetta]{kandala2017hardware}
Kandala,~A.; Mezzacapo,~A.; Temme,~K.; Takita,~M.; Brink,~M.; Chow,~J.~M.; Gambetta,~J.~M. Hardware-efficient variational quantum eigensolver for small molecules and quantum magnets. \emph{nature} \textbf{2017}, \emph{549}, 242--246\relax
\mciteBstWouldAddEndPuncttrue
\mciteSetBstMidEndSepPunct{\mcitedefaultmidpunct}
{\mcitedefaultendpunct}{\mcitedefaultseppunct}\relax
\EndOfBibitem
\bibitem[Innan \latin{et~al.}(2024)Innan, Khan, and Bennai]{innan2024quantum}
Innan,~N.; Khan,~M. A.-Z.; Bennai,~M. Quantum computing for electronic structure analysis: Ground state energy and molecular properties calculations. \emph{Materials Today Communications} \textbf{2024}, \emph{38}, 107760\relax
\mciteBstWouldAddEndPuncttrue
\mciteSetBstMidEndSepPunct{\mcitedefaultmidpunct}
{\mcitedefaultendpunct}{\mcitedefaultseppunct}\relax
\EndOfBibitem
\bibitem[Powell(1994)]{powell1994direct}
Powell,~M.~J. \emph{Advances in optimization and numerical analysis}; Springer, 1994; pp 51--67\relax
\mciteBstWouldAddEndPuncttrue
\mciteSetBstMidEndSepPunct{\mcitedefaultmidpunct}
{\mcitedefaultendpunct}{\mcitedefaultseppunct}\relax
\EndOfBibitem
\bibitem[Santagati \latin{et~al.}(2018)Santagati, Wang, Gentile, Paesani, Wiebe, McClean, Morley-Short, Shadbolt, Bonneau, Silverstone, \latin{et~al.} others]{santagati2018witnessing}
Santagati,~R.; Wang,~J.; Gentile,~A.~A.; Paesani,~S.; Wiebe,~N.; McClean,~J.~R.; Morley-Short,~S.; Shadbolt,~P.~J.; Bonneau,~D.; Silverstone,~J.~W.; others Witnessing eigenstates for quantum simulation of Hamiltonian spectra. \emph{Science advances} \textbf{2018}, \emph{4}, eaap9646\relax
\mciteBstWouldAddEndPuncttrue
\mciteSetBstMidEndSepPunct{\mcitedefaultmidpunct}
{\mcitedefaultendpunct}{\mcitedefaultseppunct}\relax
\EndOfBibitem
\bibitem[Higgott \latin{et~al.}(2019)Higgott, Wang, and Brierley]{higgott2019variational}
Higgott,~O.; Wang,~D.; Brierley,~S. Variational quantum computation of excited states. \emph{Quantum} \textbf{2019}, \emph{3}, 156\relax
\mciteBstWouldAddEndPuncttrue
\mciteSetBstMidEndSepPunct{\mcitedefaultmidpunct}
{\mcitedefaultendpunct}{\mcitedefaultseppunct}\relax
\EndOfBibitem
\bibitem[McClean \latin{et~al.}(2017)McClean, Kimchi-Schwartz, Carter, and De~Jong]{mcclean2017hybrid}
McClean,~J.~R.; Kimchi-Schwartz,~M.~E.; Carter,~J.; De~Jong,~W.~A. Hybrid quantum-classical hierarchy for mitigation of decoherence and determination of excited states. \emph{Physical Review A} \textbf{2017}, \emph{95}, 042308\relax
\mciteBstWouldAddEndPuncttrue
\mciteSetBstMidEndSepPunct{\mcitedefaultmidpunct}
{\mcitedefaultendpunct}{\mcitedefaultseppunct}\relax
\EndOfBibitem
\bibitem[Colless \latin{et~al.}(2018)Colless, Ramasesh, Dahlen, Blok, Kimchi-Schwartz, McClean, Carter, de~Jong, and Siddiqi]{colless2018computation}
Colless,~J.~I.; Ramasesh,~V.~V.; Dahlen,~D.; Blok,~M.~S.; Kimchi-Schwartz,~M.~E.; McClean,~J.~R.; Carter,~J.; de~Jong,~W.~A.; Siddiqi,~I. Computation of molecular spectra on a quantum processor with an error-resilient algorithm. \emph{Physical Review X} \textbf{2018}, \emph{8}, 011021\relax
\mciteBstWouldAddEndPuncttrue
\mciteSetBstMidEndSepPunct{\mcitedefaultmidpunct}
{\mcitedefaultendpunct}{\mcitedefaultseppunct}\relax
\EndOfBibitem
\bibitem[Fan \latin{et~al.}(2021)Fan, Liu, Li, and Yang]{fan2021equation}
Fan,~Y.; Liu,~J.; Li,~Z.; Yang,~J. Equation-of-motion theory to calculate accurate band structures with a quantum computer. \emph{The Journal of Physical Chemistry Letters} \textbf{2021}, \emph{12}, 8833--8840\relax
\mciteBstWouldAddEndPuncttrue
\mciteSetBstMidEndSepPunct{\mcitedefaultmidpunct}
{\mcitedefaultendpunct}{\mcitedefaultseppunct}\relax
\EndOfBibitem
\bibitem[Nakagawa \latin{et~al.}(2024)Nakagawa, Kamoshita, Mizukami, Sudo, and Ohnishi]{nakagawa2024adapt}
Nakagawa,~Y.~O.; Kamoshita,~M.; Mizukami,~W.; Sudo,~S.; Ohnishi,~Y.-y. Adapt-qsci: Adaptive construction of an input state for quantum-selected configuration interaction. \emph{Journal of Chemical Theory and Computation} \textbf{2024}, \emph{20}, 10817--10825\relax
\mciteBstWouldAddEndPuncttrue
\mciteSetBstMidEndSepPunct{\mcitedefaultmidpunct}
{\mcitedefaultendpunct}{\mcitedefaultseppunct}\relax
\EndOfBibitem
\bibitem[Barison \latin{et~al.}(2025)Barison, Moreno, and Motta]{barison2025quantum}
Barison,~S.; Moreno,~J.~R.; Motta,~M. Quantum-centric computation of molecular excited states with extended sample-based quantum diagonalization. \emph{Quantum Science and Technology} \textbf{2025}, \emph{10}, 025034\relax
\mciteBstWouldAddEndPuncttrue
\mciteSetBstMidEndSepPunct{\mcitedefaultmidpunct}
{\mcitedefaultendpunct}{\mcitedefaultseppunct}\relax
\EndOfBibitem
\bibitem[Huggins \latin{et~al.}(2022)Huggins, O’Gorman, Rubin, Reichman, Babbush, and Lee]{huggins2022unbiasing}
Huggins,~W.~J.; O’Gorman,~B.~A.; Rubin,~N.~C.; Reichman,~D.~R.; Babbush,~R.; Lee,~J. Unbiasing fermionic quantum Monte Carlo with a quantum computer. \emph{Nature} \textbf{2022}, \emph{603}, 416--420\relax
\mciteBstWouldAddEndPuncttrue
\mciteSetBstMidEndSepPunct{\mcitedefaultmidpunct}
{\mcitedefaultendpunct}{\mcitedefaultseppunct}\relax
\EndOfBibitem
\bibitem[Lee(2022)]{lee2022unbiasing}
Lee,~J. Unbiasing Fermionic Quantum Monte Carlo with a Quantum Computer. APS March Meeting Abstracts. 2022; pp N40--007\relax
\mciteBstWouldAddEndPuncttrue
\mciteSetBstMidEndSepPunct{\mcitedefaultmidpunct}
{\mcitedefaultendpunct}{\mcitedefaultseppunct}\relax
\EndOfBibitem
\bibitem[Nation and Treinish(2023)Nation, and Treinish]{nation2023suppressing}
Nation,~P.~D.; Treinish,~M. Suppressing quantum circuit errors due to system variability. \emph{PRX Quantum} \textbf{2023}, \emph{4}, 010327\relax
\mciteBstWouldAddEndPuncttrue
\mciteSetBstMidEndSepPunct{\mcitedefaultmidpunct}
{\mcitedefaultendpunct}{\mcitedefaultseppunct}\relax
\EndOfBibitem
\bibitem[Zhang \latin{et~al.}(2022)Zhang, Cincio, Negre, Czarnik, Coles, Anisimov, Mniszewski, Tretiak, and Dub]{zhang2022variational}
Zhang,~Y.; Cincio,~L.; Negre,~C.~F.; Czarnik,~P.; Coles,~P.~J.; Anisimov,~P.~M.; Mniszewski,~S.~M.; Tretiak,~S.; Dub,~P.~A. Variational quantum eigensolver with reduced circuit complexity. \emph{npj Quantum Information} \textbf{2022}, \emph{8}, 96\relax
\mciteBstWouldAddEndPuncttrue
\mciteSetBstMidEndSepPunct{\mcitedefaultmidpunct}
{\mcitedefaultendpunct}{\mcitedefaultseppunct}\relax
\EndOfBibitem
\bibitem[Cao \latin{et~al.}(2021)Cao, Hu, Zhang, Xu, Chen, Yu, Li, Hu, Lv, and Yung]{cao2021towards}
Cao,~C.; Hu,~J.; Zhang,~W.; Xu,~X.; Chen,~D.; Yu,~F.; Li,~J.; Hu,~H.; Lv,~D.; Yung,~M.-H. Towards a larger molecular simulation on the quantum computer: Up to 28 qubits systems accelerated by point group symmetry. \emph{arXiv preprint arXiv:2109.02110} \textbf{2021}, \relax
\mciteBstWouldAddEndPunctfalse
\mciteSetBstMidEndSepPunct{\mcitedefaultmidpunct}
{}{\mcitedefaultseppunct}\relax
\EndOfBibitem
\bibitem[Nishi \latin{et~al.}(2008)Nishi, Iwahashi, Yamane, Ouchi, Kanai, and Seki]{nishi2008electronic}
Nishi,~T.; Iwahashi,~T.; Yamane,~H.; Ouchi,~Y.; Kanai,~K.; Seki,~K. Electronic structures of ionic liquids [Cnmim]+ BF4-and [Cnmim]+ PF6-studied by ultraviolet photoemission, inverse photoemission, and near-edge X-ray absorption fine structure spectroscopies. \emph{Chemical Physics Letters} \textbf{2008}, \emph{455}, 213--217\relax
\mciteBstWouldAddEndPuncttrue
\mciteSetBstMidEndSepPunct{\mcitedefaultmidpunct}
{\mcitedefaultendpunct}{\mcitedefaultseppunct}\relax
\EndOfBibitem
\bibitem[Wrobel \latin{et~al.}(2019)Wrobel, Kubisiak, and Eilmes]{wróbel2019interactions}
Wrobel,~P.; Kubisiak,~P.; Eilmes,~A. Interactions in sodium bis (fluorosulfonyl) imide/1-ethyl-3-methylimidazolium bis (fluorosulfonyl) imide electrolytes for Na-ion batteries: insights from molecular dynamics simulations. \emph{The Journal of Physical Chemistry C} \textbf{2019}, \emph{123}, 14885--14894\relax
\mciteBstWouldAddEndPuncttrue
\mciteSetBstMidEndSepPunct{\mcitedefaultmidpunct}
{\mcitedefaultendpunct}{\mcitedefaultseppunct}\relax
\EndOfBibitem
\bibitem[Berhaut \latin{et~al.}(2019)Berhaut, Lemordant, Porion, Timperman, Schmidt, and Anouti]{berhaut2019ionic}
Berhaut,~C.~L.; Lemordant,~D.; Porion,~P.; Timperman,~L.; Schmidt,~G.; Anouti,~M. Ionic association analysis of LiTDI, LiFSI and LiPF 6 in EC/DMC for better Li-ion battery performances. \emph{RSC advances} \textbf{2019}, \emph{9}, 4599--4608\relax
\mciteBstWouldAddEndPuncttrue
\mciteSetBstMidEndSepPunct{\mcitedefaultmidpunct}
{\mcitedefaultendpunct}{\mcitedefaultseppunct}\relax
\EndOfBibitem
\bibitem[Wang \latin{et~al.}(2022)Wang, Yu, Kong, Kim, Boyle, Qin, Bao, and Cui]{wang2022liquid}
Wang,~H.; Yu,~Z.; Kong,~X.; Kim,~S.~C.; Boyle,~D.~T.; Qin,~J.; Bao,~Z.; Cui,~Y. Liquid electrolyte: The nexus of practical lithium metal batteries. \emph{Joule} \textbf{2022}, \emph{6}, 588--616\relax
\mciteBstWouldAddEndPuncttrue
\mciteSetBstMidEndSepPunct{\mcitedefaultmidpunct}
{\mcitedefaultendpunct}{\mcitedefaultseppunct}\relax
\EndOfBibitem
\end{mcitethebibliography}

\end{document}